\documentclass[aps,preprint]{revtex4-2}
\usepackage[latin9]{inputenc}
\usepackage{mathtools}
\usepackage{bm}
\usepackage{amsmath}
\usepackage{amssymb}
\usepackage{graphicx}

\makeatletter
\usepackage{amsthm}
\usepackage{amsfonts}
\usepackage{xcolor}
\usepackage{comment}
\usepackage[export]{adjustbox}
\newcommand{\R}{I\kern-0.3emR}
\newcommand{\E}{\mathbb{E}}
\makeatother

\begin{document}
\title[Societies of Neural Networks]{Legislative Rebellions and Impeachments in a Neural Network Society}
\author{Juan Neirotti}
\email{j.p.neirotti@aston.ac.uk}

\affiliation{Department of Mathematics~\\
 Aston University~\\
 The Aston Triangle, B4 7ET, Birmingham, UK}
\author{Nestor Caticha}
\email{ncaticha@usp.br}

\affiliation{Instituto de Fisica, ~\\
Universidade de Sao Paulo,~\\
 CEP 05315-970, Sao Paulo, SP, Brazil}
\date{\today}
\begin{abstract}
Inspired by studies of government overthrow in  modern presidential democracies in South
America, we present an agent based Statistical Mechanics exploration of the collective,
coordinated action of strategic political actors in the legislative chamber and the conditions 
that may result in  premature changes in the executive officeholder, such as a president's impeachment or a motion of no confidence on a prime minister. The legislative actors are information processing agents equipped with a neural network and emit opinions about issues in the presidential agenda. 
We construct a Hamiltonian, the sum of the costs for the agents for holding a set of opinions among those that could be possible. We use replica methods for two types of disorder, in the space of weights and in the network of agents' interactions. We obtain the phase diagram of the model, where the control parameters may be loosely described as indices measuring the strategic legislative support,  the presidential polling popularity and the volume of the presidential agenda under discussion. It shows an intermediate phase of coexistence of pro and against strategic behavior. This region is surrounded by pure phases where the strategic vote leans completely into the for or the against campus. 
Driven by externalities, the change of these indices may  lead the system out of  the coexistence region into the against pure phase, 
triggering a phase transition into a state  opposing the executive leader and supporting its removal from office by constitutional means. We use data from Brazil, and show the presidential trajectories that led to impeachment or not during the democratic period starting in 1989. These trajectories ended in the region of the phase diagram in accordance to the president being removed or not from office.
\end{abstract}
\keywords{Agent Based Models Statistical Mechanics Neural Networks agents, distrust,
effective polarization, Entropic Dynamics, spin glass.}
\maketitle
\newpage

\section{Introduction}

The mechanism for deposition of institutional power in several areas
of the world has changed in the last half century. A new pattern of
government overthrow took over the traditional military coup, specially
in Latin America as extensively documented in \cite{AnibalPerezLinan}.
These abrupt changes of collective behavior are typically seen in
a society which exerts pressure on the parliament to promote changes
outside the election model, by parliamentary impeachment. External
pressures which include elements such as the state of the economy,
perception of corruption or their combination may be distal reasons,
but at a closer look the correlated behavior of the parliament follows
the emboldenment that derives from the collective perception that
there is sufficient strength in the opposition camp to overthrow the
executive. Technically still within the realm of constitutional order,
despite being associated to affective rather than ideological affinity
\cite{IyengarAffective,KlarAffective_Full}, this transition mechanism
seems to bring new theoretical challenges to comparative studies of
presidentialism \cite{AnibalPerezLinan}.

To illustrate the problem and to better understand the expected characteristics
of the model, we will briefly discuss different instances were
the executive was either impeached or not by the legislative in Brazil. The reason for this is that we have access to the relevant data (\cite{Cebrap, 
Datafolha, Glauco}).
The first case corresponds to the presidency of Fernando Collor de
Mello, from Brazil. Collor won the 1989 elections in the second round
with 54\% of the votes, but his party only had 8\% of the seats in
the Chamber of Deputies and 4\% in the Senate. By March 1990, when
Collor was sworn into office and his approval ratings were at +60\%,
the consumer price index rose by 84\%;  
at this point Collor launched
his first economic plan. But, in spite of the extreme measures imposed,
government control of inflation proved elusive and popular discontent
increased. The application of a second (unsuccessful) economic plan
(Collor II), and a number of corruption cases revealed by the press
provoked a plummet on the approval ratings and triggered a number
of streets demonstrations. With very few allies in the Legislative, the
impeachment was approved by 441 votes against 38 in  the Chamber of Deputies 
and by 76 votes against 2 in the Federal Senate.

After a short transition government, F. H. Cardoso
and L. I. Lula da Silva led administrations that completed the  full term, without successful  parliamentary challenges. 
In more recent times, 
we have the presidency of Dilma
Rousseff that also ended in impeachment. Ms Rousseff's presidency began on
January 2011, with a healthy approval rating of 47\%, a balanced Senate and
majority in the Deputies' Chamber. Ms Rousseff was re-elected for a second term, however the composition of the Legislative had changed against Ms Rousseff.
Her public image was damaged by the corruption in the management of
the state-owned oil company (Petrobras) and the subsequent anti-corruption
investigation (operation \emph{Car Wash}).  The Chamber of Deputies approved the initiation of the impeachment process (367 in favor 137 against) and the Federal Senate finally removed her from office (61 in favor 20
against). 

The presidency of Jair Bolsonaro in Brazil, is another example of presidential tenure that did not end in impeachment. His support remained moderate,  in the mid 40\%. In this case a large number of impeachment attempts ($> 10^2$) were shielded from  initiating legislative
proceedings by the Legislative  Chamber leader, due to the fact that Mr Bolsonaro's polling 
was much higher than those of Collor or Rousseff and had considerable legislative support.

Certain characteristics are shared by the different examples, among them we cite:
\begin{itemize}
    \item Presidential tenures start with high approval ratings and at least some moderate legislative support.
    \item The members of the legislative chamber discuss the president's proposals
under the influence of their internal political alliances and the effective pressure exerted by the presidential approval ratings. 
\item Presidential approval ratings and legislative chamber's alliances evolve 
over time due to the emergence of items in the presidential agenda
(i.e. proposed policies) or pieces of information about the president (i.e. scandals, state of the economy).
\item There are two types of members of Congress. Those  in the first group  are quenched either in favor or opposite to the president, independently of any external influence. Their voting patterns are trivial to predict. They may be collectively called the expressive members. The second group, which we call the strategic voters,  vote in one or another direction.
\end{itemize}
    The separation into expressive and strategic voters has been widely used in Political Science with respect to voters \cite{Lnunez, Duverger, Brennan}. We borrow it to describe members of congress. In this paper we only discuss the behavior of the strategic members of parliament.
    
Furthermore, intuition permits the expectation that chances of impeachment increase under
\begin{itemize}
\item  an increase of difference in opinions between strategic legislative agents and the president,
\item  a decrease of the presidential approval rating.
\end{itemize}

But intuition may not be sufficient. There is a need to understand mechanisms for opinion formation and
alliance formation in order to provide insight for the understanding
of modern presidential democracies. Empirical evidence
coming from research in psychology supports there is a cost of dissent,
with humans trying to attain social conformity modulated by peer pressure
\cite{Asch,Schelling,Sherif,rejection} and that conformity is learned
from interactions within a social network e.g. \cite{Klucharev}.
The cost of dissension on some set of issues can be modeled using
techniques of Statistical Mechanics of disordered systems and there
is an extensive literature on polarization \cite{Lord,Baron,Vicente09,Dadenkar,Lee,Ramos,Schweitzer}
and echo chambers \cite{Gilbert2009,Zhang2,Mas,Vicario} in opinion
dynamics models. In this paper we address the collective behavior behind constitutional impeachments and legislative rebellions.
Our aim is to contribute to a twofold  discussion to this area. One is a mathematical approach based on the analysis of Complex Systems \cite{Mezard}, using replica methods and Agent Based Modeling. This leads to the identification of the pertinent effective control parameters and the order parameters  relevant to construct a phase diagram. The second is to encourage political scientists to devise empirical methods in order to measure these parameters which may better instrumentalize their approach to the study of real legislative rebellions. Before we delve into the problem, we discuss the type of result that emerges from our analysis and give a summary of what might be obtained from this approach. The discussion of how this picture emerges starts in Section \ref{Organization}.

\begin{figure}
    \includegraphics[width=\linewidth]{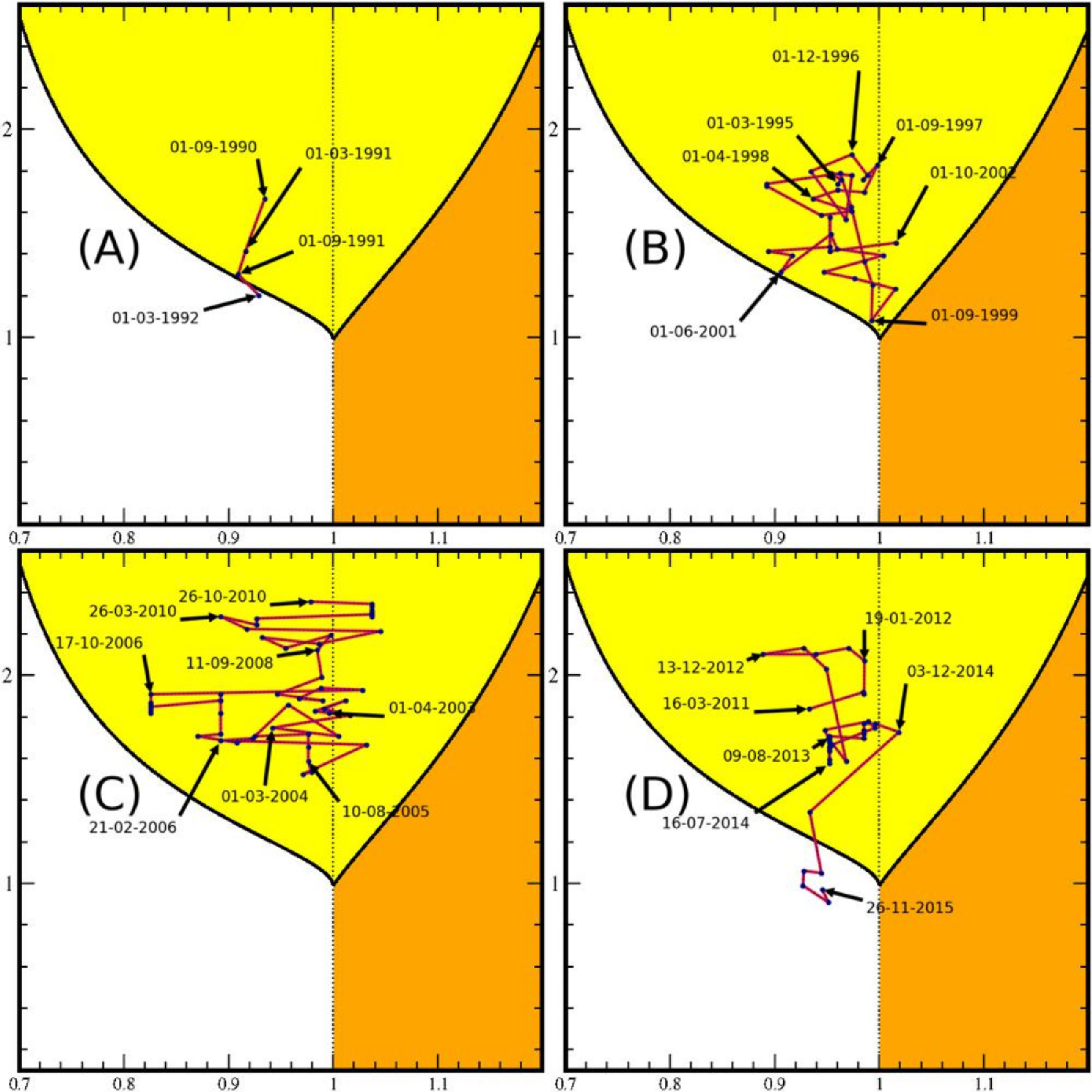}   
    \caption{(A) F. Collor de Mello was impeached, (B) F. H. Cardoso and (C) L. Lula da Silva were not impeached and (D) Dilma Rousseff was impeached.
    The x-axis is related to parliamentary support to the President, and the
    y-axis to the popular support as discussed in Section \ref{sec:trajectories}.  }
    \label{fig:trajectories}
\end{figure}

The two control parameters can be succinctly interpreted as the public approval rating  and the strategic legislative chamber ideological alignment with the president. Figure \ref{fig:trajectories} shows the phase diagram, with three regions of different collective behavior of the legislative body, as indicated by an order parameter, with the following meaning: in light gray (yellow online) the coexistence region where the strategic legislative body has a distribution of alignment with two peaks, associated to the groups for and against the executive. The dark gray (orange online) region shows a unified legislative in favor of the president, and in the white region, the president has lost support from the strategic members of congress. The lines were constructed from the data on presidential polling and from legislative voting records. The mean support of the government in legislative is measured between polling dates and only considers what are called issues of substance, where a qualified quorum is needed,  as opposed to simple procedural issues, where a win of the government position carries no significant meaning. These trajectories, which are driven by externalities,  show  for four Brazilian presidents, cases that led to constitutional legislative rebellions and impeachments and others that remained in the stereotypical democratic regime, where dialogue is still possible. In section \ref{sec:trajectories} we come back to how this figure was constructed.

\section{Organization \label{Organization}}
The model is introduced in Section \ref{sec:modelo}. The analytical
approach derives from the use of Statistical Mechanics in the space
of interactions in the Gardner style \cite{Gardner}, here applied
not only to a single perceptron agent, but to a population of such
agents, that interact on a random graph. The relevant time scales of the different changes that can
occur are discussed and this leads to the methodology appropriate
for the analysis. In Section \ref{sec:meth} we present the structure
of the agents, the relevant order parameters that characterize the
macroscopic state of the society and the analytical framework. The
two types of quenched disorder, from the issues under discussion and
the graph of interactions, lead as shown in Section \ref{sec:Saddle-point-equationsB},
to functional mean field equations that determine the thermodynamics
of the model. In Section \ref{sec:resultados} we present analytical
results obtained from the study of the saddle-point equations. Readers
interested in the lessons that can be gleaned from this toy model
should go to Section \ref{sec:politica} where the interpretation
in terms of less mathematical terms can be found. A short version
of the extensive calculations is shown in the Supplementary Material
(SM) Section in \ref{sec:SupMat}.

\section{The Model \label{sec:modelo}}
\subsection{The Hamiltonian cost}
Our strategic members-of-congress agents are simple neural networks that discuss
and emit for or against opinions about multidimensional issues. In addition there is a special agent, playing the role of the current
executive leader, called the president, also modeled by a neural network. 

The agenda,  composed by the set of $P$ substantial issues under discussion, is modeled by representing them with   $N$-dimensional binary vectors
$\bm{\xi}_{\mu}\in\{-1,+1\}^N$. A usual way to characterize its size in units of $N$ is given by $\alpha=P/N$, which is a simple measure of the agenda's complexity. 

The
$a^{\text{th}}$ agent's for or against opinion is $\sigma_{a\mu}$,
arising from the issue and its internal state ${\bf J}_{a}\in\mathbb{R}^N$, $\sigma_{a\mu}=\varphi({\bf J}_{a},\bm{\xi}_{\mu})\in\{-1,+1\}$,
where $a$ runs over the members of congress. Note the internal space of an agent, parameterized by ${\bf J}_a$ means this is far from being a simple Ising spin. 
The  president's opinion on issue $\bm{\xi}_{\mu}\in\{-1,+1\}^N$
is $\sigma_{B\mu}=\varphi({\bf B},\bm{\xi}_{\mu})$, where ${\bf
B}\in\mathbb{R}^N$
is its internal state. A specific choice of the neural networks architecture 
 $\varphi$ will
be postponed for now, but its output is binary, i.e. $\varphi\in\{-1,+1\}$.

The inner circle or clique of a particular congress-agent is represented
by an adjacency matrix $\bm{G}$ with entries $g_{ac}\ne0$ if agents
$a$ cares about the opinion of agent $c$ and zero if not. The weighed
opinion on issue $\bm{\xi}_{\mu}$ of $a$'s peers is $\Sigma_{a\mu}=\sum_{c}g_{ac}\sigma_{c\mu}$.

The model follows from simply adding costs due to different interactions that the agents have. We consider the cost for agent $a$ to hold an opinion on the $\mu$th
issue to arise from two contributions:
\begin{equation}
C_{a,\mu}=-\frac{1+\sigma_{B\mu}\sigma_{a\mu}}{2}-\frac{1-\sigma_{B\mu}\sigma_{a\mu}}{2}\sigma_{a\mu}\Sigma_{a\mu}.\label{cost0}
\end{equation}
 Equation (\ref{cost0}) implements a mechanism of \emph{corroboration}
as follows. If agent $a$ agrees with $\bf B$, i.e. $\sigma_{B\mu}\sigma_{a\mu}=1$,
only the first term contributes to the cost, which gets reduced in
one unit. If $a$ and $\bf B$ disagree, $\sigma_{B\mu}\sigma_{a\mu}=-1$, then the second term is different from zero. If the weighed opinion
of $a$'s peers is in agreement with $\bf B$, i.e. $\sigma_{B\mu}\Sigma_{a\mu}=-\sigma_{a\mu}\Sigma_{a\mu}>0$,
the cost increases, and if $\sigma_{a\mu}\Sigma_{a\mu}>0$ the cost
decreases. If agreeing with its peers is less costly than agreeing
with $\bf B$, $a$ can form a local consensus against $\bf B$, through corroboration.

A simple rearrangement of terms allows us to write the cost as:
\begin{equation}
2C_{a,\mu}=-\sigma_{B\mu}\sigma_{a\mu}-\Sigma_{a\mu}\sigma_{a\mu}+\Sigma_{a\mu}\sigma_{B\mu}-2.\label{cost}
\end{equation}
The first term describes the advantage of having the same opinion
as the president. The second, of concurring with its peers. The third
one can be attributed to the disadvantage that other members of its
peer group are in alliance with the president. The last is just an
additive constant.

The overall cost for the entire congress, defined by the microscopic
states $\{{\bf J}_{a}\}$ of the agents, the topological structure
of alliances $\bm{G}$ and the complete presidential agenda ${\cal A}=\{\bm{\xi}_{\mu},\sigma_{B\mu}\}_{\mu=1,...P}$,
gives the full Hamiltonian cost of the system, up to a constant:
\begin{eqnarray}
E(\{{\bf J}_{a}\},\mathcal{A},\bm{G})&=&\sum_{a,\mu}C_{a,\mu}= E_1+ E_2 +E_3,\label{energia} \\
E_1 &=&-\sum_{a,\mu}\sigma_{B\mu} \sigma_{a\mu} \nonumber\\
E_2 &=&-\sum_{a,\mu}\Sigma_{a\mu} \sigma_{a\mu} \nonumber\\
E_3 &=&\sum_{a,\mu}\Sigma_{a\mu} \sigma_{B\mu} \nonumber
\end{eqnarray}
with the three terms respectively describing for every agent, the interaction of the agent and the president ($E_1$), of the agent and its peers ($E_2$) and  of its peers and the president ($E_3$).
The cost for agent $a$, expression (\ref{cost}), depends on the
$g_{ac}$ in two places. In the first, which we leave as shown, it
describes the interaction of the peers $c$ with agent $a$. The second
describes the overall influence of the president on the group of peers
of $a$. To simplify matters we disregard fluctuations in the second term, and substitute $\sum_a g_{ca} \approx \langle \sum_a g_{ca} \rangle \approx\nu\eta_{0}$, where $\eta_{0}$
is the average intensity of the influence exerted by another agent
and $\nu$ the average size of the group. 
The parameter $\nu\eta_0$ represents the excess effective number of
agents with an opinion against the president's position and is a central effective parameter. Note that we don't disregard fluctuations of $\sum_a g_{ca} \sigma_{a\mu}$, expected to be larger.
Then the overall cost, up
to an additive constant, simplifies to:
\begin{equation}
E_{0}(\{{\bf J}_{a}\};\mathcal{A},\bm{G})=-\frac{1}{2}\left[(1-\nu\eta_{0})\sum_{a,\mu}\sigma_{B\mu}\sigma_{a\mu}+\sum_{\mu ac}g_{ac}\sigma_{a\mu}\sigma_{c\mu}\right].\label{energia2}
\end{equation}
\subsection{The time scales}
The choice of techniques used to analyze the system follows from a
discussion of the relevant ``physical'' time scales of the problem.
While there is no conservation law that applies to the global cost
in any strict sense, there are several relevant time scales associated
to this discussion. The agenda under discussion and the political
alliances are supposed to remain valid on a long time scale $\tau_{q}$
of around one year, certainly less than $\tau_{P}$, the 4 or 5 years
of the presidential cycle. For times of the order of $\tau_{q}$ we
expect $\nu$, the dissident clique size; $\eta_{0}$, the dissident interaction strength;  and $\alpha$, a measure of the volume
of the agenda covered, to remain constant. Agents interact and may
change opinions about the issues on a faster scale $\tau_{op}$ of
a few days. $\tau_{op}$ is the typical time elapsed between the treatment
of subsequent agenda items $\boldsymbol{\xi}_{\mu}$ and $\boldsymbol{\xi}_{\mu+1}$.
The expected value of the cost is sufficiently stable on an intermediate
time scale $\tau_{C}$, which is larger than the time scale associated
to the dynamics of the agents but much shorter than changes of the
issues of national interest, $\tau_{op}\ll\tau_{C}\ll\tau_{q}<\tau_{P}$.
$\tau_{C}$ is of the order of weeks, similar to the time validity
of presidential polls data. This separation of the time scales determines
the methodology of analysis of the problem and leads to a description
of the system with a Boltzmann distribution with a $\beta$, conjugated
to the expected value of the cost, that controls the size of the fluctuations
above the ground state. It can be interpreted as the pressure that
society at large exerts on congress. As an example, \cite{alves2016sympatric}
choose $\beta$ as a measure of the president's polling or presidential approval
rating. Since the
time scale in which the agenda and the political alliances changes
is still larger, their random effect can be averaged over as quenched
disorder. This is reasonable, since at least during $\tau_{q}$ the
prominent issues of the agenda are to some extent fixed, as are the
intra-party alliances. 

The macroscopic state of the system is characterized by order parameters
to be described below. Still within the presidential cycle, changes
due to externalities may lead to changes in the intensive parameters.
Phase transitions may occur for a congress divided into a situation
and opposition parties, to a majoritarian congress that either supports
a state of almost unanimous support for the president or it is in
almost total opposition. These transitions are to a constitutional
dictatorship regime or to a state where the conditions for impeachment
are ripe. They signal presidential crises driven by the collective
properties of congress and not by external or internal military forces
that act by simple dissolution of congress. 

\section{Methods and order parameters\label{sec:meth}}

We have not yet made explicit the \textit{a priori} measure of the
$\sigma$ variables. If it were just a product of independent uniform
measures, e.g. Ising like variables, several interesting features
of the problem would remain untouched. Thus we decided for more structured
agents, which we model by a neural network classifier with a binary
for/against, $\pm1$ output. In order to keep it analytically manageable,
we choose the simplest architecture, the single layer perceptron.
Linearly separable models, in some manner similar to the Rescorla-Wagner
model \cite{RescorlaWagner} from psychology have been shown to be
useful in describing human performance in several cases. Therefore
the dynamical variables of  agent $a$ are $N$ dimensional vectors ${\bf J}_{a}$
and its opinion on an issue is $\sigma_{a\mu}=\text{sgn}({\bf
J}_{a}\cdot\bm{\xi}_{\mu})$, where for any ${\bf V}$ and ${\bf
U}\in\mathbb{R}^N$ we have that ${\bf V}\cdot{\bf U}=\sum_{j=1}^N V_j U_j$.
The issues from the agenda are constructed by choosing independently
$P=\alpha N$ vertices of the $N$ dimensional hyper-cube with coordinates
of absolute value equal to one. $M$ is the number of congressional agents, which for the Brazilian National Congress is  $513$.

Under the assumption that the average value of $E_{0}$ in equation
\ref{energia2}, is approximately constant over a cycle of discussions
of order $\tau_{C}$ and the random agenda and alliances quenched
on the $\tau_{q}$ scale, standard arguments yield the probability
distribution of the states of the congress-agents, given by:
\begin{equation}
\mathcal{P}(\{{\bf J}_{a}\}|\beta,\mathcal{A},{\bf G})=\frac{1}{Z}\mathcal{P}_{0}(\{{\bf J}_{a}\})\exp\{-\beta E_{0}(\{{\bf J}_{a}\};\mathcal{A},{\bf G})\},\label{eq:zz}
\end{equation}
where $\mathcal{P}_{0}(\{{\bf J}_{a}\})=\prod_a(2\pi\mathrm{e})^{-N/2}\,\delta\left({\bf J}_a\cdot{\bf J}_a-N\right)$
is the \textit{a priori} measure of the agents weights, taken to be
independent and uniform over the spherical shell of radius $\sqrt{N}$
in $N$ dimensions. The discussion about the separation of time scales
require the use of quenched disorder. The macroscopic properties of
the system are obtained from the free energy $f=-\beta^{-1}\overline{\ln Z},$
averaged of the possible agendas and alliances, taken to be fixed
on the relevant time scale.

The interaction $g$ between a pair of agents, an  element of the matrix $\boldsymbol{G}$,
are assumed independent of each other and identically distributed.
They are constructed in two steps. First, the random variable $x\in \{0,1\}$ which has a Bernoulli distribution with parameter
$p$, is used to decide if there is a connection present between two
peers. Then, the strength of their interaction $\eta$ is drawn from
a Normal distribution centered at $\eta_{0}$ with variance $\Delta^{2}.$
With $\nu=2p(M-1)$ (see equation (\ref{eq:nu})), 
\begin{eqnarray}
\mathcal{P}(g,\eta,x|\nu,\eta_{0},\Delta^{2}) & =&\mathcal{P}(g|\eta,x)\mathcal{P}(x|p)\mathcal{P}(\eta|\eta_{0},\Delta^{2})\nonumber \\
 & =&\delta(g-x\eta)\left[(1-p)\delta(x)+p\delta(x-1)\right]\mathcal{N}(\eta|\eta_{0},\Delta^{2}).\label{eq:dist}
\end{eqnarray}
The problem is complicated by the existence of two sources of quenched
disorder, the agenda ${\cal A}$ and the alliances, encoded in the
matrix $\boldsymbol{G}$. An adaptation of ideas introduced in References
\cite{Bowman1982,Wong1987,Muruyama2000} to treat this problems associated
to coding theory are needed here. Due to the technical impossibility
of computing the average of a logarithm we proceed by applying the
replica formalism \cite{Mezard}, i.e.:
\begin{equation}
f=-\beta^{-1}\lim_{n\to0}\E_{{\bf \xi}, {\bf G}}\frac{Z^{n}-1}{n},\label{eq:replica}
\end{equation}
where $Z^{n}=\prod_{\gamma=1}^{n}Z^{(\gamma)}$ is the partition function
of the replicated system, each of the $n$ systems linked to a replica index $\gamma$.

Taking expectations over the alliances brings forward the following
population averages order parameters: 
\begin{equation}
\varrho_{\mu_{1}\dots\mu_{\ell}}^{\gamma_{1}\dots\gamma_{\ell}}\equiv\mathbb{E}_{{\bf \xi}, {\bf G}}\left[\frac{1}{M}\sum_{a}\left(\sigma_{a\mu_{1}}^{\gamma_{1}}\sigma_{B\mu_{1}}\dots\sigma_{a\mu_{\ell}}^{\gamma_{\ell}}\sigma_{B\mu_{\ell}}\right)\right],\label{eq:rhol}
\end{equation}
where $\varrho_{\mu}^{\gamma}$ is the average agreement of the population
with $\bf B$ on the $\mu$-th issue on the $\gamma$ replica, and $\varrho_{\mu_{1}\dots\mu_{\ell}}^{\gamma_{1}\dots\gamma_{\ell}}$
are population averages of the agreement of individual $a$ with $\bf B$
across systems $\gamma_{1}$ (on issue ${\bm \xi}_{\mu_{1}}$) to $\gamma_{\ell}$
(on issue ${\bm \xi}_{\mu_{\ell}}$). Their expectation values are $\ell$-point
correlation functions for the opinions. The introduction of these
parameters also requires the introduction of conjugate parameters
$\hat{\varrho}_{\mu_{1}\dots\mu_{\ell}}^{\gamma_{1}\dots\gamma_{\ell}}$.
Observe that $\varrho_{\mu_{1}\dots\mu_{\ell}}^{\gamma_{1}\dots\gamma_{\ell}}$
is the average of local properties, thus the conjugate variable
$\hat{\varrho}_{\mu_{1}\dots\mu_{\ell}}^{\gamma_{1}\dots\gamma_{\ell}}$
must represent the average effect of the local neighborhood on the
local agent. By imposing the replica-symmetric ansatz \cite{Bowman1982,Wong1987,Muruyama2000},
the order parameters should not present any dependency on either replica
or agenda item indexes, they should only depend on their number $\ell$.
By observing that the definition of the order parameters equation
(\ref{eq:rhol}) satisfy $-1\leq\varrho_{\mu_{1}\dots\mu_{\ell}}^{\gamma_{1}\dots\gamma_{\ell}}\leq1$,
we suppose the existence of a field $\tanh(\beta z)$ which is drawn
from two normalized distributions $\pi(z)$ and $\hat{\pi}(z)$ such
that: 
\begin{equation}
\varrho_{\mu_{1}\dots\mu_{\ell}}^{\gamma_{1}\dots\gamma_{\ell}}=\int\mathrm{d}z\,\pi(z)\tanh^{\ell}(\beta z),  \qquad\hat{\varrho}_{\mu_{1}\dots\mu_{\ell}}^{\gamma_{1}\dots\gamma_{\ell}}=\nu\int\mathrm{d}z\,\hat{\pi}(z)\tanh^{\ell}(\beta z).\label{rhos}
\end{equation}

Graph disorder introduces these two probability densities $\pi(z)$
and $\hat{\pi}(s)$, that are functional order parameters that describe
the level of consensus at the local and neighborhood levels, respectively.
It is their behavior that signals the transitions from a two parties
equilibrium to a consensus that can be either for or against the presidential
agent.

The introduction of the order  parameters in (\ref{rhos}) will be carefully discussed 
in (\ref{eq:pi1}) and (\ref{eq:pi2}).
The usual order parameters associated with the agents overlaps and
with the president are also introduced:
\begin{eqnarray}
R_{a}^{\gamma} & =&\mathbb{E}({\bf J}_{a}^{\gamma}\cdot{\bf B}/N),\quad q_{a}^{\gamma\rho}=\mathbb{E}({\bf J}_{a}^{\gamma}\cdot{\bf J}_{a}^{\rho}/N),\label{eq:q1}\\
W_{ab}^{\gamma} & =&\mathbb{E}({\bf J}_{a}^{\gamma}\cdot{\bf J}_{b}^{\gamma}/N),\quad t_{ab}^{\gamma\rho}=\mathbb{E}({\bf J}_{a}^{\gamma}\cdot{\bf J}_{b}^{\rho}/N).\label{eq:t}
\end{eqnarray}
Under the assumption of replica symmetric saddle points $R_{a}^{\gamma}=R,$
$q_{a}^{\gamma\rho}=q$, and, by Reference \cite{Neirotti14}, $W_{ab}^{\gamma}=t_{ab}^{\gamma\rho}=W$,
the properties of the system follow from the extrema of the free energy
functional (see Section \ref{sec:SupMat}): 
\begin{eqnarray}
\beta f[q,R,\pi,\hat{\pi}] & =&\mathop{\mathrm{extr}}_{q,R,\pi,\hat{\pi}}\left\{ -\frac{1}{2}\left(\ln(1-q)+\frac{q-R^{2}}{1-q}\right)+\right.\nonumber \\
 && +\alpha\nu\int\mathrm{d}z\,\mathrm{d}s\,\pi(z)\hat{\pi}(s)\,\ln\left(\frac{1+\tanh(\beta s)\tanh(\beta z)}{1-\tanh(\beta s)}\right)-\nonumber \\
 && -\alpha\frac{\nu}{2}\int\mathrm{d}z_{1}\,\mathrm{d}z_{2}\,\pi(z_{1})\pi(z_{2})\,\left\langle \ln\left[1+\tanh(\beta\eta)\tanh(\beta z_{1})\,\tanh(\beta z_{2})\right]\right\rangle _{\eta}-\nonumber \\
 && \left.-\alpha\left\langle \left\langle \ln\left[1+\epsilon^{-1}\mathcal{H}\left(-\sqrt{\frac{q}{1-q}}x\right)\right]\right\rangle _{x}\right\rangle _{y}\right\} ,\label{eq:freenergyB}
\end{eqnarray}
where the averages are taken over the following random variables
$\eta\sim\mathcal{N}(\eta|\eta_{0},\Delta^{2}),$ $y\sim\mathsf{P}(y|\hat{\pi})
$, and $x\sim2\mathcal{N}(x|0,1)\mathcal{H}\left(-Rx/\sqrt{q-R^{2}}\right)$,
where $\mathcal{H}(t)$ is the Gardner error function $\mathcal{H}(t)=\int_{t}^{\infty}\mathrm{d}x\mathcal{N}(x|0,1)$.
We used the short hand $\epsilon=\epsilon(\beta,\nu\eta,y)=\left[\exp(2\beta(1-\nu\eta_{0}+y))-1\right]^{-1}$.
The free energy is a functional of the normalized distributions $\pi$
and $\hat{\pi}$. The new variable $y$'s distribution is:
\begin{equation}
\mathsf{P}(y|\hat{\pi})\equiv\int\frac{\mathrm{d}\hat{y}}{2\pi}\mathrm{e}^{-iy\hat{y}}\exp\left[\nu\left(\int\mathrm{d}s\hat{\pi}(s)\mathrm{e}^{i\hat{y}s}-1\right)\right].\label{eq:pp}
\end{equation}
The characteristic function $\phi_{s}(\hat{y})$ of $\hat{\pi}(s)$,
and the generator function of the cumulants of $s$, $K_{s}(\hat{y})$
are:
\begin{eqnarray}
\phi_{s}(\hat{y}) & =&\int\mathrm{d}s\hat{\pi}(s)\mathrm{e}^{i\hat{y}s},\label{eq:phis}\\
K_{s}(\hat{y}) & =&\log\phi_{s}(\hat{y}).\label{eq:logphis}
\end{eqnarray}
 We observe that there exists a random variable $u$, with
a generator of cumulants function given by $K_{u}(\hat{y})\equiv\phi_{s}(\hat{y})-1$.
Add $\nu$ independent copies of $u$ to define $y=\sum_{i=1}^{\nu}u_{i}$,
since:
\begin{eqnarray}
\mathsf{P}(y|\hat{\pi}) & =&\int\frac{\mathrm{d}\hat{y}}{2\pi}\mathrm{e}^{-iy\hat{y}}\exp\left[\nu K_{u}(\hat{y})\right],\label{eq:pp2}\\
 & =&\int\frac{\mathrm{d}\hat{y}}{2\pi}\mathrm{e}^{-iy\hat{y}}\left[\phi_{u}(\hat{y})\right]^{\nu},\label{eq:pp3}
\end{eqnarray}
where the $u_{i}$ are random variables with the property that the
$r$th cumulant of $u$ is equal to $\E(s^{r}|\hat{\pi})$, the $r$th
moment of $s$. Hence as $\nu$ grows, $y$ becomes normal. The $r$th
order cumulant $\kappa_{r}^{(y)}$ of $y$ satisfies:
\begin{equation}
\kappa_{r}^{(y)}=\nu\kappa_{r}^{(u)}=\nu\mathbb{E}(s^{r}|\hat{\pi}),\label{eq:kk}
\end{equation}
 so the cumulants of $y$ are constructed by accumulation of the cumulants
of $u$ or of the moments of $s$. It automatically follows that:
\begin{eqnarray}
\mathbb{E}(y|\mathsf{P}) & =&\nu\mathbb{E}(s|\hat{\pi})\label{Uno}\\
\mathbb{E}(y^{2}|\mathsf{P})-\mathbb{E}(y|\mathsf{P})^{2} & =&\nu\mathbb{E}(s^{2}|\hat{\pi}).\label{Dos}
\end{eqnarray}
Since $\E(s^{2}|\hat{\pi})$ turns out to be proportional to $\eta_{0}^{2}$,
$y$'s variance is proportional to $1/\nu$ in the relevant region
where $\nu\eta_{0}$ is of order $1.$ 

\section{Saddle point equations\label{sec:Saddle-point-equationsB}}

The extreme of the free energy (\ref{eq:freenergyB}) is determined
by the saddle point equations, which determine the order parameters
in a self-consistent way. The distribution $\hat{\pi}(s)$ satisfies:
\begin{equation}
\hat{\pi}(s)=\int\mathrm{d}z\int\mathrm{d}y\,\mathsf{P}(z,s,y|\hat{\pi})=\int\mathrm{d}z\int\mathrm{d}y\,\mathsf{P}(y|\hat{\pi})\mathsf{P}(z|y)\mathsf{P}(s|z),\label{eq:pp-1}
\end{equation}
where
\begin{eqnarray}
\mathsf{P}(z|y) & =&\left\langle \delta\left[z-\beta^{-1}g(x;\epsilon,q)\right]\right\rangle _{x}\label{eq:xx}\\
\mathsf{P}(s|z) & =&\left\langle \delta\left(s-\beta^{-1}\mathrm{arctanh}\left[\tanh(\beta\eta)\tanh(\beta z)\right]\right)\right\rangle _{\eta},
\end{eqnarray}
 and
\begin{equation}
g(x;\epsilon,q)=\frac{1}{2}\ln\frac{1+\epsilon}{\epsilon}+\frac{1}{2}\ln\frac{1-\mathcal{H}_{+}}{\mathcal{H}_{+}};\text{with}\text{\,\,}\mathcal{H}_{+}=\mathcal{H}\left({\displaystyle \sqrt{\frac{q}{1-q}}}x\right).\label{eq:ggg}
\end{equation}

The equations for $\pi(z)$ and $\hat{\pi}(s)$ are: 
\begin{eqnarray}
\pi(z) & =&\int\mathrm{d}y\,\mathsf{P}(y|\hat{\pi})\left\langle \delta\left[z-\beta^{-1}g(x;\epsilon,q)\right]\right\rangle _{x},\label{eq:sp1}\\
\hat{\pi}(s) & =&\int\mathrm{d}z\,\pi(z)\left\langle \delta\left(s-\beta^{-1}\,\mathrm{arctanh}\left[\tanh(\beta\eta)\tanh(\beta z)\right]\right)\right\rangle _{\eta}.\label{eq:sp2}
\end{eqnarray}
This shows that $\pi(z)$ is the distribution of the local field $z$
associated to the agent, that is constructed over the influence of
its neighborhood through $\mathsf{P}(y|\hat{\pi})$, the distribution
of consensus in the neighborhood of the agent, and the influence of
the agenda through the average over $x$. These two contributions
represent the sources the agent uses to form its opinion. The distribution
of the neighborhood effective field $\hat{\pi}(s)$ acting on the
local agent is obtained by averaging over the distribution of the
local agent field through $\pi(z)$ and through the distribution of
influences through the average over $\eta$. Observe that if there
is agreement between agent and president and if the influence between
peers is strong (large $\eta$), the neighborhood field $s$ becomes
large and positive. If the agent does not give any importance to its
peers ($\eta=0$), the distribution $\hat{\pi}(s)$ becomes a delta
function centered at zero, and the system decouples. 

In addition, these functional saddle point equations also depend on
the usual parameters $q$ and $R$, which satisfy 
\begin{eqnarray}
\frac{q-R^{2}}{1-q} & =&\frac{\alpha}{\pi}\int\mathrm{d}y\,\mathsf{P}(y|\hat{\pi})\int\mathcal{D}x\,\frac{\exp\left(-{\displaystyle \frac{qx^{2}}{1-q}}\right)\,\mathcal{H}\left(-{\displaystyle \frac{Rx}{\sqrt{q-R^{2}}}}\right)}{\left[\epsilon+\mathcal{H}\left(-{\displaystyle \sqrt{\frac{q}{1-q}}}x\right)\right]^{2}}\label{eq:sd3}\\
\frac{R}{\sqrt{1-q}} & =&\frac{\alpha}{\pi}\sqrt{\frac{q}{q-R^{2}}}\int\mathrm{d}y\,\mathsf{P}(y|\hat{\pi})\int\mathcal{D}x\frac{\exp\left\{ -{\displaystyle \left(\frac{q}{1-q}+\frac{R^{2}}{q-R^{2}}\right)\frac{x^{2}}{2}}\right\} }{\epsilon+\mathcal{H}\left(-{\displaystyle \sqrt{\frac{q}{1-q}}}x\right)},\label{eq:sp4}
\end{eqnarray}

The numerical solution of this set of equations is discussed in Section
\ref{sec:Map}.

\section{macroscopic characterization of the model\label{sec:resultados}}

In Section \ref{sec:Map} we demonstrate that for sufficiently large
neighborhoods ($\nu>O(1)$), for sufficiently high presidential approval $\beta$,
and for a very narrow distribution of social strengths, i.e. $\Delta\ll\eta_{0}$
and $\mathcal{P}(\eta)=\mathcal{N}(\eta|\eta_{0},\Delta^{2})$, there
are three possible solutions for equations (\ref{eq:sp1}) and (\ref{eq:sp2}).
Two of them are the pure \emph{supportive} state, in the sense that supports the presidential {\it status quo}, obtained if $\nu\eta_{0}<1$
and the other is the \emph{opposition} pure state if $\nu\eta_{0}>1$.
There is a possibility of a third solution which is a mixture of the
two pure states, that appears in the region of the phase space where
dialogue between opposite positions may exist.  Defining a parameter
$\Lambda$ as: 
\begin{equation}
\Lambda(R)\equiv\frac{\mathrm{sgn}(R)}{2\beta}\frac{q}{1-q},\label{Lambda}
\end{equation}
permits plotting a partial phase diagram, presented in figure
\ref{mapa-2}. $|\Lambda(R)|^{-1}$ plays the role of an effective presidential approval rating. In the region $(|\Lambda|,\eta_{0})\in\mathbb{A}$
a convex combination of both pure states is found. For $(|\Lambda|,\eta_{0})\notin\mathbb{A}$
the distributions can be expressed as: 
\begin{eqnarray}
\hat{\pi}_{0}(z) & \equiv&\mathcal{N}\left(z\left|\mathcal{I}_{0}^{\star}(\Lambda,\eta_{0}),\eta_{0}^{2}-[\mathcal{I}_{0}^{\star}(\Lambda,\eta_{0})]^{2}+\Delta^{2}\right.\right)\label{eq:c2}\\
\pi_{0}(s) & \equiv&\mathcal{N}\left(s\left|1+\nu[\mathcal{I}_{0}^{\star}(\Lambda,\eta_{0})-\eta_{0}]+\frac{1}{2}\Lambda,\nu(\eta_{0}^{2}+\Delta^{2})+\frac{3}{4}\Lambda^{2}\right.\right)\label{eq:c3}\\
\mathsf{P}_{0}(y|\hat{\pi}) & \equiv&\mathcal{N}\left(y\left|\nu\mathcal{I}_{0}^{\star}(\Lambda,\eta_{0}),\nu(\eta_{0}^{2}+\Delta^{2})\right.\right),\label{eq:c4}
\end{eqnarray}
where $\mathcal{I}_{0}^{\star}$ is the only solution to
\begin{equation}
\mathcal{I}_{0}^{\star}=\eta_{0}\mathrm{erf}\left(\frac{1-\nu\eta_{0}+\nu\mathcal{I}_{0}^{\star}+\frac{1}{2}\Lambda(R)}{\sqrt{2\left[\nu(\eta_{0}^{2}+\Delta^{2})+\frac{3}{4}\Lambda(R)^{2}\right]}}\right)\label{eq:map-1}
\end{equation}
outside region $\mathbb{A}$ (this equation is developed in Section
\ref{sec:Map}, equation (\ref{eq:map})). We have observed that in
the region of interest $\nu\eta_{0}\sim O(1),$ the variance of $\mathsf{P}_{0}(y|\hat{\pi})$
is of order $O(\nu^{-1})$, therefore we can approximate this distribution
by: 
\begin{equation}
\mathsf{P}_{0}(y|\hat{\pi})\approx\delta(y-\nu\mathcal{I}_{0}^{\star}).\label{eq:c4-1}
\end{equation}
Inside the region $\mathbb{A}$ we have mixed states described by:
\begin{eqnarray}
\hat{\pi}_{\mathrm{m}}(z) & \equiv& h_{+}\mathcal{N}\left(s\left|\mathcal{I}_{+}^{\star},\Delta^{2}\right.\right)+h_{-}\mathcal{N}\left(s\left|\mathcal{I}_{-}^{\star},\Delta^{2}\right.\right)\label{eq:c2-1}\\
\pi_{\mathrm{m}}(s) & \equiv&\mathcal{N}\left(s\left|1+\nu[\mathcal{I}^{\star}-\eta_{0}]+\frac{1}{2}\Lambda,\nu(\eta_{0}^{2}+\Delta^{2})+\frac{3}{4}\Lambda^{2}\right.\right)\label{eq:c3-1}\\
\mathsf{P}_{\mathrm{m}}(y|\hat{\pi}) & \equiv&\mathcal{N}\left(y\left|\nu\mathcal{I}^{\star},\nu\{\left\langle (\mathcal{I}^{\star})^{2}\right\rangle +\Delta^{2}\}\right.\right),\label{eq:c4-2}
\end{eqnarray}
where $\mathcal{I}_{\pm}^{\star}$ are the stable solutions to equation
(\ref{eq:map-1}) in $\mathbb{A}$, $h_{\pm}$ are suitable weights
(\ref{eq:hache}) satisfying $0\leq h_{\pm}\leq1$ and $h_{+}+h_{-}=1$,
$\mathcal{I}^{\star}$ is the mixed solution: 
\begin{eqnarray}
\mathcal{I}^{\star} & \coloneqq& h_{+}\mathcal{I}_{+}^{\star}+h_{-}\mathcal{I}_{-}^{\star}\label{eq:medmix}\\
\left\langle (\mathcal{I}^{\star})^{2}\right\rangle  & \coloneqq& h_{+}(\mathcal{I}_{+}^{\star})^{2}+h_{-}(\mathcal{I}_{-}^{\star})^{2},\label{eq:varmix}
\end{eqnarray}
and, given that for all $(\lambda,\eta_{0})\in\mathbb{A}$, $\nu\eta_{0}\sim O(1),$
we have that: 
\begin{equation}
\mathsf{P}_{\mathrm{m}}(y|\hat{\pi})\approx\delta\left(y-\nu\mathcal{I}^{\star}\right).\label{eq:c43}
\end{equation}

The application of equations (\ref{eq:c4-1}) and (\ref{eq:c43})
into equations (\ref{eq:sd3}) and (\ref{eq:sp4}) produce the following
expressions 
\begin{eqnarray}
\frac{q-R^{2}}{1-q} & =&\frac{\alpha}{\pi}\left(\mathrm{e}^{\kappa}-1\right)^{2}\int\mathcal{D}x\,\frac{\exp\left(-{\displaystyle \frac{qx^{2}}{1-q}}\right)\,\mathcal{H}\left(-{\displaystyle \frac{Rx}{\sqrt{q-R^{2}}}}\right)}{\left[{\displaystyle 1}+\left(\mathrm{e}^{\kappa}-1\right)\mathcal{H}\left(-{\displaystyle \sqrt{\frac{q}{1-q}}}x\right)\right]^{2}}\label{eq:qR-1}\\
\frac{R}{\sqrt{1-q}} & =&\frac{\alpha}{\pi}\left(\mathrm{e}^{\kappa}-1\right)\sqrt{\frac{q}{q-R^{2}}}\int\mathcal{D}x\frac{\exp\left\{ -{\displaystyle \left(\frac{q}{1-q}+\frac{R^{2}}{q-R^{2}}\right)\frac{x^{2}}{2}}\right\} }{{\displaystyle 1}+\left(\mathrm{e}^{\kappa}-1\right)\mathcal{H}\left(-{\displaystyle \sqrt{\frac{q}{1-q}}}x\right)},\label{eq:R-1}
\end{eqnarray}
where $\kappa\equiv2\beta(1-\nu\eta_{0}+\nu\mathcal{I^{\star}}).$
Observe that these equations are invariant under the following transformation
$(\kappa,q,R)\to(-\kappa,q,-R)$.

At very high presidential approval $\beta$, the equations (\ref{eq:sd3}) and
(\ref{eq:sp4}) can be expressed as: 
\begin{eqnarray}
\frac{q_{\pm}-R_{\pm}^{2}}{1-q_{\pm}} & =&\frac{\alpha}{\pi}\int\mathcal{D}x\,\frac{\exp\left(-{\displaystyle \frac{q_{\pm}x^{2}}{1-q_{\pm}}}\right)\,\mathcal{H}\left(-{\displaystyle \frac{R_{\pm}x}{\sqrt{q_{\pm}-R_{\pm}^{2}}}}\right)}{\left[\mathcal{H}\left(-{\displaystyle \sqrt{\frac{q_{\pm}}{1-q_{\pm}}}}x\right)\right]^{2}}\label{eq:sd3-1}\\
\frac{R_{\pm}}{\sqrt{1-q_{\pm}}} & =&\pm\frac{\alpha}{\pi}\sqrt{\frac{q_{\pm}}{q_{\pm}-R_{\pm}^{2}}}\int\mathcal{D}x\frac{\exp\left\{ -{\displaystyle \left(\frac{q_{\pm}}{1-q_{\pm}}+\frac{R_{\pm}^{2}}{q_{\pm}-R_{\pm}^{2}}\right)\frac{x^{2}}{2}}\right\} }{\mathcal{H}\left(-{\displaystyle \sqrt{\frac{q_{\pm}}{1-q_{\pm}}}}x\right)},\label{eq:sp4-1}
\end{eqnarray}
where the sub-index $+(-)$ is valid for $\nu\eta_{0}<(>)1.$ The
$\beta\to\infty$ solutions satisfy $q_{\pm}=\pm R_{\pm}$. These
results justify naming the solution with sub-index $+$ as {\it in favor}
and the solution with sub-index $-$ as {\it in opposition}. Similar behavior
is observed for finite but large values of the presidential approval.

For sufficiently large presidential approval ratings, sufficiently large $\nu$ and a
volume of information $\alpha\gg\beta,$ we can also demonstrate that:
\begin{eqnarray}
q & =&1-\frac{Q_{0}^{2}}{\alpha^{2}}+o(\alpha^{-2})\label{eq:qass}\\
Q_{0} & =&\frac{(2\pi)^{3/2}}{2+\sqrt{\pi}}\label{eq:qo}
\end{eqnarray}
and 
\begin{equation}
R=\begin{cases}
q+2\pi\sqrt{3}Q_{0}^{3}\alpha^{-3}\mathrm{e}^{-2\beta} & \nu\eta_{0}<1\\
-q-2\pi\sqrt{3}Q_{0}^{3}\alpha^{-3}\mathrm{e}^{-2\beta(\nu\eta_{0}-1)} & 1<\nu\eta_{0}
\end{cases}.\label{eq:rmas}
\end{equation}

Due to the odd parity of $R(\kappa)$, we can conclude that the plane
$(\nu\eta_{0},\beta^{-1})$ is divided into two phases, the in favor
phase for which $R>0$ and $\nu\eta_{0}<1$ and a opposition phase
with $R<0$ and $\nu\eta_{0}>1.$

Consider the set of points with coordinates ${(\beta^{-1},\nu\eta=1,\alpha)}$
such that the parameter defined in equation (\ref{Lambda}) becomes
$|\Lambda^{\star}|=0.411$ (see figure \ref{mapa-2}). In consequence,
the solution of equation (\ref{eq:map-1}) over this line is $\nu\mathcal{I}^{\star}=0.988$
and the correspondent index $\kappa$ becomes a function of the overlap
$q$, i.e. 
\begin{equation}
\kappa^{\star}(q)=\frac{\nu\mathcal{I}^{\star}}{|\Lambda^{\star}|}\frac{q}{1-q}.\label{kappastar}
\end{equation}

By solving the equations (\ref{eq:qR-1}) and (\ref{eq:R-1}) with
$\kappa$ given by (\ref{kappastar}), we obtain the curve presented
in figure \ref{temperatura}.

\begin{figure}
\begin{centering}
\includegraphics[scale=0.5]{fig2} 
\par\end{centering}
\caption{Critical presidential approval against size of agenda. When the number of substantial 
issues ($P =\alpha N$) in the legislature is not sufficiently large, there
is always a phase around the opinion boundary $\nu\eta=1$ where opposition
and supportive states coexist. This area represents the collection
of points $(\alpha,\beta^{-1})$ where a discussion between members
of the chamber with different positions may occur. There is a critical
value of the agenda's size $\alpha^{\star}=1.534(1)$ bellow which there
is always room for discussion, no matter how low the presidential approval $\beta$
is. Above this threshold there is always a minimum presidential approval 
$\beta(\alpha)$
such that above it there is no more discussion and positions are definitely
set.}
\label{temperatura} 
\end{figure}
\begin{figure}
\begin{centering}
\includegraphics[scale=0.5]{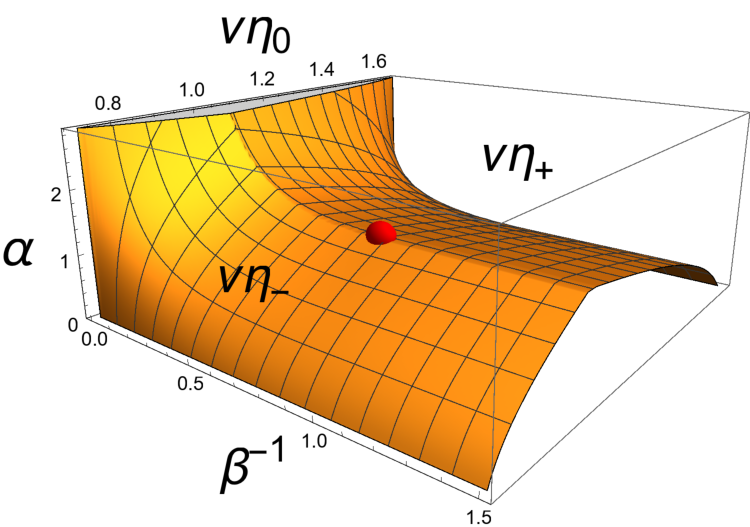} 
\par\end{centering}
\caption{Phase diagram of the system in terms of the parameters $\nu\eta_{0}$,
$\beta^{-1}$ and $\alpha$. There are two phases separated by the
plane $\nu\eta_{0}=1$. For $\nu\eta_{0}<1$ we have that $R>0$ and
the average consensus is in favor of $\bf B$. For $\nu\eta_{0}>1$, $R<0$
and the average position of the agents is to form local alliances
against the president $\bf B$. In all the points of the space above the
surfaces $\nu\eta_{+}$ and $\nu\eta_{-}$, the distribution describing
the position of the neighborhood, given by $\hat{\pi}(z)$, is sharply
picked at $+\eta_{0}$, for the supportive position, i.e. $\nu\eta_{0}<1$,
or at $-\eta_{0}$, for the opposition position, i.e. $\nu\eta_{0}>1$.
In the region bellow the surfaces $\nu\eta_{+}$ and $\nu\eta_{-}$
we have the same phase separation at $\nu\eta_{0}=1$ but the contribution
from the neighborhood is a mixture of a opposition component plus a
supportive component. The circle at coordinates $\nu\eta_{0}=1$,
$\beta^{-1}=0.824$ and $\alpha=1.534$ is the critical point presented
in figure \ref{temperatura}. The phase diagram presented in figure
\ref{mapa} has been obtained by cutting sections at constant $\alpha$
from this three-dimensional plot, and the the red sphere corresponds
to the first value of $\alpha$ ($=\alpha^{\star}$) for which the
behavior presented in figure \ref{mapa} c) is observed.}
\label{mapa-5} 
\end{figure}

We solve for the properties of the equilibrium state valid in the
$\tau_{q}$ time scale in the intensive parameters of the system:
$\beta$ the presidential approval, $\alpha$ a measure of the complexity of the
agenda and $\nu\eta_{0}$ a measure of the peer pressure by other
agents in congress which arises from the mean number of interlocutors
$\nu$ and the mean intensity of their interaction $\eta_{0}$. These
results are presented as the phase diagrams shown in figure \ref{mapa-5}.

\begin{figure}
\begin{centering}
\includegraphics[scale=0.65]{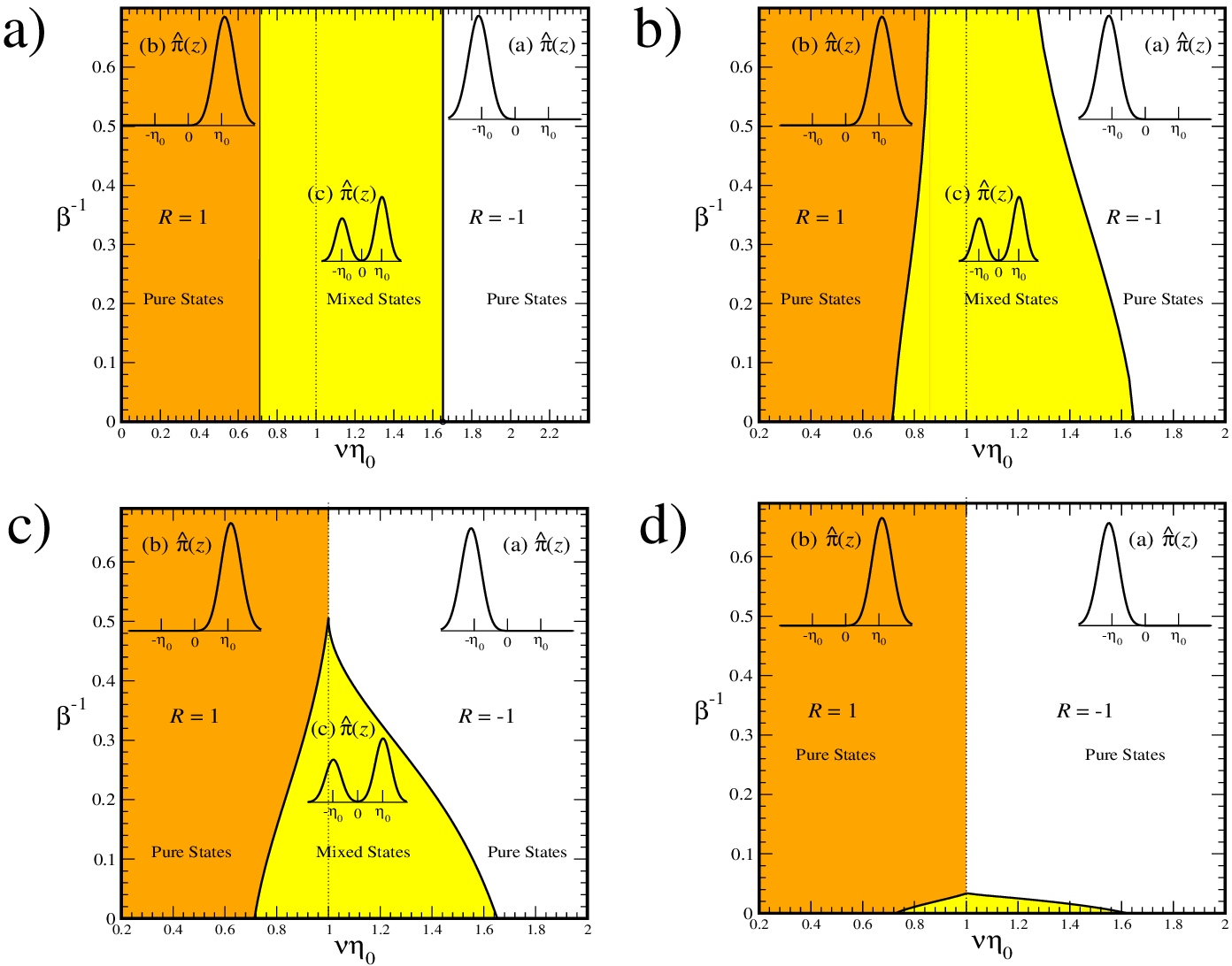} 
\par\end{centering}
\caption{Phase diagram of the system for different volumes of information $\alpha.$
For $\nu\eta_{0}<1$ and all values of $\beta$ and $\alpha$, the
average agreement with the executive $R$ is positive and the state
of the system is mostly supportive, i.e. the local consensus, described
by the distribution $\hat{\pi}$ is centered at $\eta_{0}>0$. For
$\nu\eta_{0}>1$ and all values of $\beta$ and $\alpha$, the level
of agreement with $\bf B$ is negative and the local consensus is against
the president. For values of $\alpha<\alpha^{\star}\approx1.534$
there is always an area around the boundary $\nu\eta_{0}=1,$ represented
in light gray (yellow on-line), where mixed states {[}equations (\ref{eq:c2-1}),
(\ref{eq:c3-1}), and (\ref{eq:c4-2}){]} develop, whereas outside
this area we have pure states only {[}given by equations (\ref{eq:c2}),
(\ref{eq:c3}), and (\ref{eq:c4}){]}. For values of $\alpha\protect\geq\alpha^{\star}$
the area where the mixed states develop gets a mixed-triangular shape,
with vertexes at $(\beta^{-1}=0,\nu\eta_{0}=1.651(1)),$ $(\beta^{-1}=0,\nu\eta_{0}=0.717(1)),$
and $(\beta=\beta(\alpha),\nu\eta_{0}=1).$ In particular we observe
that $\beta^{\star}\equiv\beta(\alpha^{\star})\approx1.214$ and for
all $\alpha^{\star}<\alpha'<\alpha,$ $\beta(\alpha)>\beta(\alpha')>\beta(\alpha^{\star}).$
These results correspond to fixed values of $\nu=10$ and $\Delta=0.01.$
The area in white corresponds to pure opposition states and the area
in dark gray (orange on-line) correspond to supportive pure states.
The panels a), b), c) and d) correspond to $\alpha=0$, $\alpha\lesssim\alpha^{\star}$,
$\alpha\gtrsim\alpha^{\star}$ and $\alpha\gg\alpha^{\star}$ respectively.\label{mapa}}
\end{figure}

By fixing the value of $\alpha$, we can study the behavior of the
system for a given volume of information. We constructed figure \ref{mapa}
by solving equation (\ref{eq:map-1}) for different values of $\beta^{-1}$
and $\nu\eta_{0}$ at fixed values of $\alpha$ with $\nu=10$ and
$\Delta=0.01$. The full lines separate pure-state areas {[}in white,
for $R<0$ and in dark gray (orange on-line) for $R>0$, given by
equations (\ref{eq:c2}), (\ref{eq:c3}), and (\ref{eq:c4}){]} from
mixed-state areas {[}in gray (yellow on-line), given by equations
(\ref{eq:c2-1}), (\ref{eq:c3-1}), and (\ref{eq:c4-2}){]}. We also
found that for values of $\alpha<\alpha^{\star}=1.534(1)$ the mixed
states are contained into a mixed-triangular-shaped area, with vertexes
at $(\beta^{-1}=0,\nu\eta_{0}=1.651(1)),$ $(\beta^{-1}=0,\nu\eta_{0}=0.717(1)),$
and $(\beta=\beta(\alpha),\nu\eta_{0}=1).$ In particular we observe
that $\beta^{\star}\equiv\beta(\alpha^{\star})\approx1.214(1)$ and
for all $\alpha^{\star}<\alpha'<\alpha,$ $\beta(\alpha)>\beta(\alpha')>\beta(\alpha^{\star}).$
The lightly shaded (yellow on-line) region close to the boundary ($\nu\eta_{0}=1$)
is characterized by a mixture of states that represents a state of
dialogue, where the influence on the agents from their neighborhoods
come from both sides of the argument. The larger the complexity of
the agenda ($\alpha$) the smaller the size of this region.

To complete our analysis and for very low values of $\beta$ we obtain
the following values for the parameters $R$ and $q$: 
\begin{eqnarray}
q & \approx&\frac{2\alpha}{\pi}\beta^{2}(1-\nu\eta_{0})^{2}\left(\frac{2\alpha}{\pi}+1\right)\label{eq:sd3-2}\\
R & \approx&\frac{2\alpha}{\pi}\beta(1-\nu\eta_{0})[1-2\beta(1-\nu\eta_{0})].\label{eq:sp4-2}
\end{eqnarray}

\section{Constructing the trajectories\label{sec:trajectories}}
We now turn to the confrontation of the model and empirical data obtained from legislative voting records \cite{Cebrap} and presidential polling data \cite{Datafolha}. There are several details  \cite{Glauco} of the voting data that have to take into account in order to illustrate the history of events during a presidential tenure. These will be the subject of a forthcoming paper written from the perspective of political science analysis. For the current purposes, of describing collective behavior in a complex system, we do a simple analysis and proceed as follows. There are 25 types of votes in the Legislative Chamber of Brazil, which can be grouped roughly into Substantial and Procedural. Some require simple majority to pass, while others require some type of qualified quorum, such as half plus one, or 3/5 of the total number of members of parliament. We selected votes  only of the Substantial  and qualified quorum types. 

The frequency of Presidential polls changes, but they roughly occurs every few months. Call the times of occurrence $T_n$. For each time interval $(T_{n-1},T_n]$ between polls we  measured the mean of votes for and against the government positions. Its  difference defines a variable $M_n$ used to represent government support. 

Popular support is measured by the approval rate. The results are given as the percentage in three categories: ``optimal/good'', ``regular'' and ``bad/terrible''. The approval $A_n$ index is constructed by adding the ``optimal/good'' plus half of  ``regular'' at time $T_n$. 

The empirical trajectories are obtained by joining the set of points $\{M_n,A_n\}$. The problem now is to place these paths on the phase diagram, i.e making them trajectories in the thermodynamic space.  
Two issues have to be considered. First, which phase diagram? These amounts to choosing the statistical mechanics variables adequate to discuss the thermodynamics state. 

What is the analog in the statistical mechanics formulation of the presidential approval? A natural choice is to use the  temperature $\beta^{-1}$. This seems natural since the Lagrange multiplier controls the magnitude of the fluctuations above the ground state of the cost function. For large temperatures,  large fluctuations about the minima are possible and for large popular approval,  dissent from peers will not carry a large  political weight. But
 the diagram then depends on $q$, which itself depends on $\alpha$, see figure \ref{mapa}. However in terms of the scaling $\tilde \beta = \frac{\Lambda^*}{|\Lambda|}\propto \beta (1-q)/q$, the diagram collapses and can be represented in two dimensional  space. These simplifies matters, since no empirical determination of $q$ is needed. Of course this points into possible future directions for a more detailed analysis. 
For the other coordinate we use the inverse of $\nu\eta_0$, which describes the effective coupling between agents. 

The second issue deals with how to relate the 
 empirical and theoretical pairs. We use linear transformations between empirical and theoretical variables. The parameters of the transformations are chosen such that Cardoso's trajectory remains in the mixed (dialogue) phase. Despite two very serious crises, when the government stability was in peril, he remained in office. The other three trajectories are then fixed and shown in figure \ref{fig:trajectories}. Collor and Rousseff end up in the region with no support form the strategic representatives, and Lula da Silva is easily within the mixed phase, never being in danger of an impeachment. The transformations are 
 \begin{eqnarray}
    A  &=& a_A \tilde \beta +b_A \nonumber\\
    M &=& a_M \frac{1}{\nu\eta_0} +b_M.
 \end{eqnarray}
Figure \ref{fig:trajectories} uses $[a_A,b_A, a_M, b_M] =[50,27,2000,-1790] $. These are not fundamental constants, since changes to the model or to the selection of voting procedures would change their value. The problem of determining the relevant transformations, i.e. the adequate ``Boltzmann constants'', remains open.

\section{Discussion \label{sec:discussion}}

During the last few  decades \cite{Galam08, 
Castellano} the application of Statistical
Mechanics techniques to model social problems have produced a number
of interesting results, not only providing new insights about social phenomena but also showing predictive capabilities \cite{Galam2017}.
Inspired by these ideas, we introduced a model for the phenomenon
of impeachment in presidential democracies. The political agents, who have a simple neural network that enables making decisions on issues, interact with an external meta-agent $\bf B$,
which represents the executive, and with peers in the legislative
chamber. The model balances the need to include psychological characteristics of
observed behavior \cite{Asch, Kumar}, the complexity of the social interactions
\cite{Schweitzer,Liu}, and the analytical tractability of the mathematical
expressions constructed.

For sufficiently large number of alliances
$\nu>O(1),$ the saddle point equations (\ref{eq:sp1} to \ref{eq:sp4})
can be solved in pairs. The first two (\ref{eq:sp1}) and (\ref{eq:sp2})
involving the distributions $\pi$ and $\hat{\pi}$ connected with
the distribution of alliances and the pair (\ref{eq:sd3}) and (\ref{eq:sp4}),
connected to the parameters associated with the discussion of the
presidential agenda. The solution to (\ref{eq:sp1}) and (\ref{eq:sp2})
has been expressed using the parameter $\Lambda$ defined in equation
(\ref{Lambda}), which brings an input from the disorder-from-learning
part of the problem into the disorder-from-graph part of the problem.
In a similar manner, the parameter $\kappa$ that helps to express
the solution of equations (\ref{eq:sd3}) and (\ref{eq:sp4}), introduces
effects from the disorder-from-graph part of the problem into the
disorder-from-learning part of the problem. The constraints that emerged
from expressing the solution in terms of these parameters have helped
constructing the phase diagram presented in figure \ref{mapa-2}.

We obtained a set of sensible results 
for a graph with an average of $\nu$ links per vertex (the number
of dissidents). In this setting and considering a steady president,
i.e. $\bf B$ constant, we found that there exist two possible pure positions.
One characterized by an overall average attitude in favor of the president
characterized by a positive and increasing (with the volume of information)
average agreement $R$, that we dubbed the presidential supportive state, and
other with a negative and decreasing value of $R,$ the opposition
state. These states are also characterized by a sharp picked distribution
of neighbors' influences $\hat{\pi}$, centered at $\mathcal{I}_{+}^{\star}$
($\mathcal{I}_{-}^{\star}$) for the supportive (opposition) state.
From figure \ref{mapa} we also showed that for volumes of information
bellow a critical value $\alpha^{\star}=1.534(1),$ there is a region
in the plane $(\nu\eta_{0},\beta^{-1})$, in a form of a band around
$\nu\eta_{0}=1,$ where mixed states, defined by the equations (\ref{eq:c2-1}),
(\ref{eq:c3-1}), and (\ref{eq:c4-2}), exist. The mixture is explicit
in equation (\ref{eq:c2-1}) that presents the influence on an agent
by its neighborhood ($\hat{\pi}$) as a combination of the two sides
of the argument. This band of mixed states collapses into a triangle
with vertexes at $(\beta^{-1}=0,\nu\eta_{0}=1.651(1)),$ $(\beta^{-}1=0,\nu\eta_{0}=0.71845(1)),$
and $(\beta=\beta(\alpha),\nu\eta_{0}=1),$ for values of $\alpha>\alpha^{\star}.$
We also observed that the larger the volume of information the smaller
the triangle area, i.e. $\alpha>\alpha'$ implies that $\beta(\alpha)>\beta(\alpha').$
The interpretation of this behavior is as follows: when information
is limited (low $\alpha$) and for values of effective dissidents
$\nu\eta_{0}\simeq1$ the influence from the neighborhood to the agent
is formed by a combination of positions in favor and against the executive.
In this region the overlap $R,$ which represents the average agreement
with $\bf B$ still has a well defined sign given by $\mathrm{sgn}(R)=\mathrm{sgn}(1-\nu\eta_{0})$,
but is the result from two pure-state contributions. In this region
coexist the two positions, in pro and against the executive $\bf B$.
Definite positions are not set yet, thus propitiating a state of dialogue.
The more information is fed to the system the smaller this region
becomes. There is a critical value of information $\alpha^{\star}=1.534(1),$
beyond which this behavior is only observed for presidential approval ratings
lower than
$\beta^{\star}=1.214(1).$ In other words, the more information is
provided the purer the contribution to the agents opinion from their
neighborhoods and the lower the chances for a dialogue between opposed
positions. For very large values of $\alpha$ and $\nu\eta_{0}\approx1$
coexistence exists only if the presidential approval $\beta$ is sufficiently high.
Thus, only a president with high index of popularity can guarantee
a discussion of the topics in the agenda between opposite positions
of the legislative. 

Under the light of the cases used as motivation for our model we observe
that there are events, represented by particular items of the executive's
agenda, that are so momentous in the formation of opinions that can
be considered critical (e.g. Collor de Mello's economic plans, Rousseff's Petrobras scandal, Cardoso' s September 1999 crisis, 2001 energy crisis), to the point that, immediately after they occur,
opposite positions in pro or against the executive's proposals become
more consolidated, and the influence of the neighborhood on the agents
becomes more polarized (on either position) and the dialogue-prone
region gets reduced. If the public rejects the proposals, $\beta$
diminishes and the executive may find itself in front of a polarized
legislative chamber that either supports it  or not. If the negative information instances persist
and neither the public or the chamber supports the president, the
executive may find itself facing an impeachment procedure. Also observe that in
order to be impeached a president must find itself facing an adverse legislative
chamber ($\nu\eta_0>1$) outside the dialogue-prone phase (coexistence phase). 

Rousseff started her tenure with high values of approval (high $\beta$) and a
favorable composition of the legislative ($\nu\eta_0<1$). Over time  and for a
sufficiently large value of $\alpha$ the dialogue-prone coexistence phase in the legislative
got reduced, the approval rate ($\beta$) decreased and the composition
of the chamber became adverse ($\nu\eta_0>1$), leaving the state of Ms 
Rousseff's presidency in the side of the adverse pure states of the chamber
and open for an impeachment process. 

This type of minimal model   is capable of capturing some of the complex relationships that link the political powers and the dynamics of social opinions because it takes into account the double random quenched disorder of agenda and alliances. The analytical treatment is possible by assuming that its general properties are robust to a number of approximations which have been used in other contexts dealing with quenched disorder. The replica symmetry  remains unbroken. Moreover, the inter-agent
interactions considered are such that only mutual reciprocity between agents is
allowed, $\eta_{ab}=\eta_{ba}$, no equilibrium solution could be reached
otherwise.
 
Several theoretical natural extensions to this work can be foreseen. Other connectivity graphs could be considered at the expense of making the calculations much harder.   The evolution of opinions in the presence of an adaptive
social rule that slowly changes following the average position of
the population was studied in  \cite{Neirotti18}. In the present context, this would translate into allowing a changing $\bf B$. As a consequence, the contribution from socially neutral issues \cite{Healy} could become a factor, 
as it can be observed by the presence
of the parameter $W$, see equation \ref{eq:t}, which represents the overlap between the representation
of different agents (and it is a measure of the level of agreement
between them). We expect that, if a similar setting is imposed in
the present framework, the free energy functional should
be dependent also on a parameter $W$, revealing the contribution
from the socially neutral issues to the system.

Also and more importantly, the control of the agenda allows strategies that can change the political scenario. Issues at the borders of doubt (i.e.
$\boldsymbol{\xi}_{0}$ such that ${\bf B}\cdot\boldsymbol{\xi}_{0}=0$ or ${\bf J}\cdot\boldsymbol{\xi}_{0}=0$) are
interesting since they lead to exponentially fast learning \cite{Kinouchi92}  in the online teacher-student scenario and may provide a recipe to avoid the dynamical traps that lead to quenched or extreme
polarization.
\section{Conclusion \label{sec:politica}}

The characterization of the macroscopic  state of the system is a first step in
the construction of a framework  to understand parliamentary rebellions that
might lead or not to impeachments.  In figure \ref{fig:trajectories} we show 
the evolution in the space of  compound variables $\tilde \beta$ and $(\nu
\eta_0)^{-1}$.  The phase diagrams can be loosely collapsed using $\tilde
\beta=\beta/\beta_c(\alpha)$, obtained by scaling $\beta$ with the critical
value $\beta_c(\alpha)$ presented in figure \ref{temperatura}. The exact scaling
is obtained from the effective presidential approval $|\Lambda|^{-1}$, defined in equation
\ref{Lambda}.   The strategic agents effective value $\nu \eta_0$ arises from
the loyal base and opposition relative numbers and a coupling constant
describing the peer pressure influence agents exert on each other. As
externalities work to produce an external pressure and influence the size of the
agenda under discussion, the system evolves in the phase diagram and it can
cross phase boundaries into the impeachment regime. 

 An important characteristic of Statistical Mechanics is that it can identify aggregate variables, which if known, would characterize macroscopic behavior. Hence, their identification and recognition as relevant, may induce the development of empirical methodology in order to produce estimates of the parameters, that would in future models, allow predictions and not only description of what has occurred.

\textbf{Acknowledgment:} We thank Glauco Peres da Silva for discussions on the Brazilian legislative data used in this paper.  This work received partial support from CNAIPS-NAP
USP. 

\section{Supplementary Material}

\subsection{Calculation of the averages over ${\cal A},$ ${\bf B}$ and ${\boldsymbol{G}}$\label{sec:SupMat}}

By observing that $\mathcal{P}(g_{ac})=\int\mathrm{d}\eta_{ac}\mathcal{N}(\eta_{ac}|\eta_{0},\Delta^{2})\sum_{x_{ac}=0,1}[p\delta_{x_{ac},1}+(1-p)\delta_{x_{ac},0}]\,\delta(g_{ac}-x_{ac}\eta_{ac})$,
where the Kronecker's delta is $\delta_{X,Y}=1$ if $X=Y$ and 0 otherwise
and Dirac's delta is $\int_{\Omega}\mathrm{d}x\delta(x-x_{0})=1$
if $x_{0}\in\Omega$ and 0 otherwise, the replicated partition function
is: 
\begin{eqnarray}
\overline{Z^{n}}(\beta) & \equiv & \int\mathrm{d}{\bf{B}}\mathcal{P}({\bf{B}})\int\prod_{\mu}\mathrm{d}\boldsymbol{\xi}_{\mu}\mathcal{P}(\boldsymbol{\xi}_{\mu})\prod_{a}\prod_{c}\int\mathrm{d}g_{ac}\mathcal{P}(g_{ac})\int\prod_{\gamma=1}^{n}\prod_{a}\mathrm{d}{\bf{J}}_{a}^{\gamma}\mathcal{P}({\bf{J}}_{a}^{\gamma})\nonumber \\
 &  & \prod_{\gamma\mu a}\exp\left\{ \beta\sum_{c\in\mathbb{N}_{a}}x_{ac}\eta_{ac}\mathrm{sgn}\left(\frac{{\bf{J}}_{a}^{\gamma}\cdot\boldsymbol{\xi}_{\mu}}{\sqrt{N}}\right)\mathrm{sgn}\left(\frac{{\bf{J}}_{c}^{\gamma}\cdot\boldsymbol{\xi}_{\mu}}{\sqrt{N}}\right)\right\} \nonumber \\
 &  & \prod_{\gamma\mu a}\exp\left\{ (1-\nu\eta_{0})\beta\mathrm{sgn}\left(\frac{{\bf{J}}_{a}^{\gamma}\cdot\boldsymbol{\xi}_{\mu}}{\sqrt{N}}\right)\mathrm{sgn}\left(\frac{{\bf{B}}\cdot\boldsymbol{\xi}_{\mu}}{\sqrt{N}}\right)\right\} ,\label{eq:z1}
\end{eqnarray}
and by defining the variables: 
\begin{equation}
\lambda_{a,\mu}^{\gamma}\equiv\frac{{\bf{J}}_{a}^{\gamma}\cdot\boldsymbol{\xi}_{\mu}}{\sqrt{N}},\qquad u_{\mu}\equiv\frac{{\bf{B}}\cdot\boldsymbol{\xi}_{\mu}}{\sqrt{N}}\label{eq:vars}
\end{equation}
and, by defining the overlaps: 
\begin{eqnarray}
R_{a}^{\gamma}\equiv\frac{{\bf{J}}_{a}^{\gamma}\cdot{\bf{B}}}{N}, & \qquad & W_{ab}^{\gamma}\equiv\frac{{\bf{J}}_{a}^{\gamma}\cdot{\bf{J}}_{b}^{\gamma}}{N},\nonumber \\
q_{a}^{\gamma\rho}\equiv\frac{{\bf{J}}_{a}^{\gamma}\cdot{\bf{J}}_{a}^{\rho}}{N}, &  & t_{ab}^{\gamma\rho}\equiv\frac{{\bf{J}}_{a}^{\gamma}\cdot{\bf{J}}_{b}^{\rho}}{N},\label{eq:params}
\end{eqnarray}
we have that the expectation over patterns is: 
\begin{eqnarray}
\left\langle \cdot\right\rangle _{{\cal A}} & \equiv & \int\prod_{\mu}\mathrm{d}\boldsymbol{\xi}_{\mu}\mathcal{P}(\boldsymbol{\xi}_{\mu})\exp\left(i\sum_{\gamma\mu a}\hat{\lambda}_{a\mu}^{\gamma}\frac{{\bf{J}}_{a}^{\gamma}\cdot\boldsymbol{\xi}_{\mu}}{\sqrt{N}}+i\sum_{\mu}\hat{u}_{\mu}\frac{{\bf{B}}\cdot\boldsymbol{\xi}_{\mu}}{\sqrt{N}}\right)\nonumber \\
 & = & \int\prod_{\gamma a}\frac{\mathrm{d}R_{a}^{\gamma}\mathrm{d}\hat{R}_{a}^{\gamma}}{2\pi/N}\,\exp\left(i\sum_{\gamma a}\hat{R}_{a}^{\gamma}(NR_{a}^{\gamma}-{\bf{J}}_{a}^{\gamma}\cdot{\bf{B}})\right)\nonumber \\
 &  & \int\prod_{\gamma}\prod_{a<b}\frac{\mathrm{d}W_{ab}^{\gamma}\mathrm{d}\hat{W}_{ab}^{\gamma}}{2\pi/N}\,\exp\left(i\sum_{\gamma}\sum_{a<b}\hat{W}_{ab}^{\gamma}(NW_{ab}^{\gamma}-{\bf{J}}_{a}^{\gamma}\cdot{\bf{J}}_{b}^{\gamma})\right)\nonumber \\
 &  & \int\prod_{a}\prod_{\gamma<\rho}\frac{\mathrm{d}q_{a}^{\gamma\rho}\mathrm{d}\hat{q}_{a}^{\gamma\rho}}{2\pi/N}\,\exp\left(i\sum_{a}\sum_{\gamma<\rho}\hat{q}_{a}^{\gamma\rho}(Nq_{a}^{\gamma\rho}-{\bf{J}}_{a}^{\gamma}\cdot{\bf{J}}_{a}^{\rho})\right)\nonumber \\
 &  & \int\prod_{\gamma<\rho}\prod_{a<b}\frac{\mathrm{d}t_{ab}^{\gamma\rho}\mathrm{d}\hat{t}_{ab}^{\gamma\rho}}{2\pi/N}\,\exp\left(i\sum_{a<b}\sum_{\gamma<\rho}\hat{t}_{ab}^{\gamma\rho}(Nt_{ab}^{\gamma\rho}-{\bf{J}}_{a}^{\gamma}\cdot{\bf{J}}_{b}^{\rho})\right)\nonumber \\
 &  & \exp\left\{ -\frac{1}{2}\sum_{\mu}\left[\sum_{\gamma a}\left(\hat{\lambda}_{a\mu}^{\gamma}\right)^{2}+2\sum_{\gamma a}\sum_{\gamma<\rho}\hat{\lambda}_{a\mu}^{\gamma}\hat{\lambda}_{a\mu}^{\rho}q_{a}^{\gamma\rho}+2\sum_{\gamma a}\sum_{a<b}\hat{\lambda}_{a\mu}^{\gamma}\hat{\lambda}_{b\mu}^{\gamma}W_{ab}^{\gamma}+\right.\right.\nonumber \\
 &  & \left.\left.+2\sum_{\gamma a}\sum_{\gamma<\rho}\sum_{a<b}\hat{\lambda}_{a\mu}^{\gamma}\hat{\lambda}_{b\mu}^{\rho}t_{ab}^{\gamma\rho}+2\sum_{\gamma a}\hat{u}_{\mu}\hat{\lambda}_{a\mu}^{\gamma}R_{a}^{\gamma}+\hat{u}_{\mu}^{2}\right]\right\} +O(N^{-1}).\label{eq:m2}
\end{eqnarray}
By considering the distribution of the synaptic vector ${\bf B}$
as $\mathcal{P}(\boldsymbol{B})=\prod_{k}\delta(B_{k}-1)$ and by
defining the matrices: 
\begin{eqnarray}
[\hat{\boldsymbol{Q}}]_{a,b}^{\gamma,\rho} & \equiv & i\left\{ \delta^{\gamma,\rho}\left(\delta_{a,b}\hat{\ell}_{a}^{\gamma}+(1-\delta_{a,b})\hat{W}_{a,b}^{\gamma}\right)+(1-\delta^{\gamma,\rho})\left(\delta_{a,b}\hat{q}_{a}^{\gamma,\rho}+(1-\delta_{a,b})\hat{t}_{a,b}^{\gamma,\rho}\right)\right\} \label{eq:qhat}
\end{eqnarray}
\begin{eqnarray}
[\boldsymbol{Q}]_{a,b}^{\gamma,\rho} & \equiv & \delta^{\gamma,\rho}\left(\delta_{a,b}+(1-\delta_{a,b})W_{a,b}^{\gamma}\right)+(1-\delta^{\gamma,\rho})\left(\delta_{a,b}q_{a}^{\gamma,\rho}+(1-\delta_{a,b})t_{a,b}^{\gamma,\rho}\right)\label{eq:q}
\end{eqnarray}
we have that the average over synaptic vectors become: 
\begin{eqnarray}
\left\langle \cdot\right\rangle _{{\bf B},\{\bf{J}_{a}^{\gamma}\}} & = & \int\prod_{\gamma,a}\frac{\mathrm{d}\hat{\ell}_{a}^{\gamma}}{4\pi}\exp\left(i\frac{N}{2}\sum_{\gamma,a}\hat{\ell}_{a}^{\gamma}-N\ln|\hat{\boldsymbol{Q}}|-\frac{1}{2}\sum_{a,b}\sum_{\gamma,\rho}\hat{R}_{a}^{\gamma}[\hat{\boldsymbol{Q}}^{-1}]_{a,b}^{\gamma,\rho}\hat{R}_{b}^{\rho}-\frac{nNM}{2}\right),\label{eq:jb}
\end{eqnarray}
which renders the following expression for the partition function:
\begin{eqnarray}
\overline{Z^{n}}(\beta) & = & \int\prod_{\gamma a}\frac{\mathrm{d}\hat{\ell}_{a}^{\gamma}}{4\pi}\int\prod_{\gamma a}\frac{\mathrm{d}R_{a}^{\gamma}\mathrm{d}\hat{R}_{a}^{\gamma}}{2\pi/N}\,\int\prod_{\gamma}\prod_{a<b}\frac{\mathrm{d}W_{ab}^{\gamma}\mathrm{d}\hat{W}_{ab}^{\gamma}}{2\pi/N}\int\prod_{a}\prod_{\gamma<\rho}\frac{\mathrm{d}q_{a}^{\gamma\rho}\mathrm{d}\hat{q}_{a}^{\gamma\rho}}{2\pi/N}\int\prod_{\gamma<\rho}\prod_{ab}\frac{\mathrm{d}t_{ab}^{\gamma\rho}\mathrm{d}\hat{t}_{ab}^{\gamma\rho}}{2\pi/N}\nonumber \\
 &  & \exp\left(\frac{N}{2}\mathrm{tr}\boldsymbol{Q}\hat{\boldsymbol{Q}}-\frac{N}{2}\ln|\hat{\boldsymbol{Q}}|-\frac{N}{2}\sum_{ab}\sum_{\gamma,\rho}\hat{R}_{a}^{\gamma}\left[\hat{\boldsymbol{Q}}^{-1}\right]_{ab}^{\gamma\rho}\hat{R}_{b}^{\rho}+iN\sum_{\gamma a}\hat{R}_{a}^{\gamma}R_{a}^{\gamma}-\frac{nNM}{2}\right)\nonumber \\
 &  & \int\prod_{\gamma\mu a}\frac{\mathrm{d}\lambda_{a\mu}^{\gamma}\mathrm{d}\hat{\lambda}_{a\mu}^{\gamma}}{2\pi}\exp\left(-i\sum_{\gamma\mu a}\hat{\lambda}_{a\mu}^{\gamma}\lambda_{a\mu}^{\gamma}\right)\nonumber \\
 &  & \int\prod_{\mu}\mathcal{D}u_{\mu}\exp\left(i\sum_{\gamma\mu a}\hat{\lambda}_{a\mu}^{\gamma}R_{a}^{\gamma}u_{\mu}+(1-\nu\eta_{0})\beta\sum_{\gamma\mu a}\mathrm{sgn}(\lambda_{a\mu}^{\gamma}u_{\mu})\right)\nonumber \\
 &  & \exp\left\{ -\frac{1}{2}\sum_{\mu}\left[\sum_{\gamma a}\left[1-(R_{a}^{\gamma})^{2}\right]\left(\hat{\lambda}_{a\mu}^{\gamma}\right)^{2}+2\sum_{\gamma a}\sum_{\gamma<\rho}\left[q_{a}^{\gamma\rho}-R_{a}^{\gamma}R_{a}^{\rho}\right]\hat{\lambda}_{a\mu}^{\gamma}\hat{\lambda}_{a\mu}^{\rho}+\right.\right.\nonumber \\
 &  & \left.\left.+2\sum_{\gamma a}\sum_{a<b}\left[W_{ab}^{\gamma}-R_{a}^{\gamma}R_{b}^{\gamma}\right]\hat{\lambda}_{a\mu}^{\gamma}\hat{\lambda}_{b\mu}^{\gamma}+2\sum_{\gamma a}\sum_{\gamma<\rho}\sum_{b}\left[t_{ab}^{\gamma\rho}-R_{a}^{\gamma}R_{b}^{\rho}\right]\hat{\lambda}_{a\mu}^{\gamma}\hat{\lambda}_{b\mu}^{\rho}\right]\right\} \nonumber \\
 &  & \left\langle \exp\left\{ \beta\sum_{\gamma\mu a}\sum_{a\neq c}x_{ac}\eta_{ac}\mathrm{sgn}(\lambda_{a\mu}^{\gamma}\lambda_{c\mu}^{\gamma})\right\} \right\rangle _{\!\!\!{\boldsymbol{G}}}+O(N^{-1})\label{eq:z2}
\end{eqnarray}
in the limit of large $N$ we find that 
\begin{eqnarray}
\hat{R}_{a}^{\gamma} & = & i\sum_{\rho,b}[\hat{\boldsymbol{Q}}]_{a,b}^{\gamma,\rho}R_{b}^{\rho}\label{eq:sol1}\\
\left[\hat{\boldsymbol{Q}}^{-1}\right]_{a,b}^{\gamma,\rho} & = & [\boldsymbol{K}]_{a,b}^{\gamma,\rho}\equiv[\boldsymbol{Q}]_{a,b}^{\gamma,\rho}-R_{a}^{\gamma}R_{b}^{\rho}\label{eq:sol2}
\end{eqnarray}
and to express the extraction of the asymptotic behavior of integrals
of the form $I_{N}\equiv\int_{x_{1}}^{x_{2}}\mathrm{d}x\,\mathrm{e}^{-Ng(x)}$
in the limit $N\to\infty$ through Laplace's method, we denote: $\mathop{\mathrm{extr}}_{x}I_{N}\equiv\mathrm{e}^{-Ng(x_{0})+O(\log N)}$
where $x_{0}$ is such that $g(x_{0})\leq g(x)$ for all $x\in[x_{1},x_{2}]$,
so we can write: 
\begin{eqnarray}
\overline{Z^{n}}(\beta) & = & \mathop{\mathrm{extr}}_{\boldsymbol{K}}\left\{ \exp\left(\frac{N}{2}\ln|\boldsymbol{K}|\right)\right.\nonumber \\
 &  & \int\prod_{\gamma\mu a}\frac{\mathrm{d}\lambda_{a\mu}^{\gamma}\mathrm{d}\hat{\lambda}_{a\mu}^{\gamma}}{2\pi}\exp\left(-i\sum_{\gamma\mu a}\hat{\lambda}_{a\mu}^{\gamma}\lambda_{a\mu}^{\gamma}\right)\nonumber \\
 &  & \int\prod_{\mu}\mathcal{D}u_{\mu}\,\exp\left(i\sum_{\gamma a}\hat{\lambda}_{a\mu}^{\gamma}R_{a}^{\gamma}u_{\mu}+(1-\nu\eta_{0})\beta\sum_{\gamma\mu a}\mathrm{sgn}(\lambda_{a\mu}^{\gamma}u_{\mu})\right)\nonumber \\
 &  & \exp\left[-\frac{1}{2}\sum_{\gamma\mu a}\left[1-(R_{a}^{\gamma})^{2}\right]\left(\hat{\lambda}_{a\mu}^{\gamma}\right)^{2}-\sum_{\gamma\mu a}\sum_{\gamma<\rho}\left[q_{a}^{\gamma\rho}-R_{a}^{\gamma}R_{a}^{\rho}\right]\hat{\lambda}_{a\mu}^{\gamma}\hat{\lambda}_{a\mu}^{\rho}-\right.\nonumber \\
 &  & \left.-\sum_{\gamma\mu a}\sum_{a<b}\left[W_{ab}^{\gamma}-R_{a}^{\gamma}R_{b}^{\gamma}\right]\hat{\lambda}_{a\mu}^{\gamma}\hat{\lambda}_{b\mu}^{\gamma}-\sum_{\gamma\mu a}\sum_{\gamma<\rho}\sum_{b}\left[t_{ab}^{\gamma\rho}-R_{a}^{\gamma}R_{b}^{\rho}\right]\hat{\lambda}_{a\mu}^{\gamma}\hat{\lambda}_{b\mu}^{\rho}\right]\nonumber \\
 &  & \left.\left\langle \exp\left\{ \beta\sum_{\gamma\mu a}\sum_{a\neq c}x_{ac}\eta_{ac}\mathrm{sgn}(\lambda_{a\mu}^{\gamma})\mathrm{sgn}(\lambda_{c\mu}^{\gamma})\right\} \right\rangle _{\!\!\!\boldsymbol{G}}\right\} ,\label{eq:z3}
\end{eqnarray}
where $\mathcal{D}x\equiv(2\pi)^{-1/2}\mathrm{d}x\,\exp(-x^{2}/2)$.
Also, by imposing the Replica Symmetric (RS) ansatz: $R_{a}^{\gamma}=R,$
$q_{a}^{\gamma,\rho}=q,$ and by following \cite{Neirotti14} we
can assume that $t_{ab}^{\gamma\rho}=W_{ab}^{\gamma}=W,$ then by
defining 
\begin{equation}
\mathcal{C}_{a\mu}\equiv\frac{\sqrt{W-R^{2}}y_{\mu}+Ru_{\mu}+\sqrt{q-W}y_{a\mu}}{\sqrt{1-q}}\label{eq:c}
\end{equation}
and by observing that the logarithm of the matrix $\boldsymbol{K}$
in the RS approach is 
\begin{equation}
\ln|\boldsymbol{K}|=nM\left(\ln(1-q)+\frac{q-R^{2}}{1-q}\right)+O(n^{2}),\label{eq:matriz k}
\end{equation}
we have that, after the integration over the variables $\{\hat{\lambda}_{a\mu}^{\gamma}\},$
the partition function becomes:

\begin{eqnarray}
\overline{Z^{n}}(\beta) & = & \mathop{\mathrm{extr}}_{R,q,W}\left\{ \exp\left[\frac{nNM}{2}\left(\ln(1-q)+\frac{q-R^{2}}{1-q}\right)\right]\right.\nonumber \\
 &  & \int\prod_{\mu}\mathcal{D}u_{\mu}\prod_{\mu}\mathcal{D}y_{\mu}\prod_{\mu a}\mathcal{D}y_{a\mu}\prod_{\gamma\mu a}\frac{\mathrm{d}\lambda_{a\mu}^{\gamma}}{\sqrt{2\pi}}\nonumber \\
 &  & \exp\left[-\frac{1}{2}\sum_{\gamma\mu a}(\lambda_{a\mu}^{\gamma}-\mathcal{C}_{a\mu})^{2}+(1-\nu\eta_{0})\beta\sum_{\gamma\mu a}\mathrm{sgn}\left(\lambda_{a\mu}^{\gamma}u_{\mu}\right)\right]\nonumber \\
 &  & \left.\left\langle \exp\left\{ \beta\sum_{\gamma\mu a}\sum_{a\neq c}x_{ac}\eta_{ac}\mathrm{sgn}(\lambda_{a\mu}^{\gamma})\mathrm{sgn}(\lambda_{c\mu}^{\gamma})\right\} \right\rangle _{\!\!\!\boldsymbol{G}}\right\} .\label{eq:z4}
\end{eqnarray}
The average over the graph variables is: 
\begin{eqnarray}
\Upsilon & \equiv&\left\langle \exp\left\{ \beta\sum_{\gamma\mu a}\sum_{a\neq c}x_{ac}\eta_{ac}\mathrm{sgn}(\lambda_{a\mu}^{\gamma})\mathrm{sgn}(\lambda_{c\mu}^{\gamma})\right\} \right\rangle _{\!\!\!\boldsymbol{G}}\nonumber \\
 & =&\int\prod_{ac}\frac{\mathrm{d}\eta_{ac}}{\sqrt{2\pi\Delta^{2}}}\exp\left[-\frac{(\eta_{ac}-\eta_{0})^{2}}{2\Delta^{2}}\right]\prod_{ac}\left\{ 1-p+p\,\prod_{\gamma\mu}\exp\left[\beta\mathrm{sgn}(\lambda_{a\mu}^{\gamma})\mathrm{sgn}(\lambda_{c\mu}^{\gamma})\right]\right\} \nonumber \\
 & =&(1-p)^{M(M-1)}\int\prod_{ac}\frac{\mathrm{d}\eta_{ac}}{\sqrt{2\pi\Delta^{2}}}\exp\left[-\frac{(\eta_{ac}-\eta_{0})^{2}}{2\Delta^{2}}\right]\nonumber \\
 && \prod_{ac}\left\{ 1+\frac{p}{1-p}\,\prod_{\gamma\mu}\left[\cosh(\beta\eta_{ac})+\mathrm{sgn}(\lambda_{a\mu}^{\gamma}\lambda_{c\mu}^{\gamma})\sinh(\beta\eta_{ac})\right]\right\} \nonumber \\
 && \quad\prod_{ac}\left\{ 1+\frac{p}{1-p}\,\cosh(\beta\eta_{ac})^{nP}\prod_{\gamma\mu}\left[1+\tanh(\beta\eta_{ac})\mathrm{sgn}\left(\lambda_{a\mu}^{\gamma}\lambda_{c\mu}^{\gamma}\right)\right]\right\} .\label{eq:upsi}
\end{eqnarray}
Observe that we are assuming that the number of neighbors, in average,
must be $\nu\ll M:$ 
\begin{eqnarray}
\sum_{a}x_{ac}\mathcal{P}(x_{ac}) & =&p(M-1)=\frac{\nu}{2}\label{eq:nu}
\end{eqnarray}
\begin{eqnarray}
\Upsilon & =&\left(1-\frac{\nu}{2(M-1)}\right)^{M(M-1)}\int\prod_{ac}\frac{\mathrm{d}\eta_{ac}}{\sqrt{2\pi\Delta^{2}}}\exp\left[-\frac{(\eta_{ac}-\eta_{0})^{2}}{2\Delta^{2}}\right]\nonumber \\
 & &\prod_{ac}\left\{ 1+\frac{p}{1-p}\,\cosh(\beta\eta_{ac})^{nP}\left[1+\tanh(\beta\eta_{ac})\sum_{\gamma\mu}\mathrm{sgn}\left(\lambda_{a\mu}^{\gamma}\lambda_{c\mu}^{\gamma}\right)\right.\right.+\nonumber \\
 & &\qquad\qquad\left.\left.+\tanh(\beta\eta_{ac})^{2}\sum_{\langle\gamma_{1}\mu_{1};\gamma_{2}\mu_{2}\rangle}\mathrm{sgn}\left(\lambda_{a\mu_{1}}^{\gamma_{1}}\lambda_{c\mu_{1}}^{\gamma_{1}}\right)\mathrm{sgn}\left(\lambda_{a\mu_{2}}^{\gamma_{2}}\lambda_{c\mu_{2}}^{\gamma_{2}}\right)+\dots\right]\right\} \nonumber \\
 & =&\exp\left\{ -\frac{\nu}{2}M\right.+\nonumber \\
 & &+\left.\frac{\nu}{2}M\sum_{\ell=0}^{nP}\left\langle \cosh(\beta\eta)^{nP}\tanh(\beta\eta)^{\ell}\right\rangle _{\eta}\sum_{\langle\gamma_{1}\mu_{1};\dots;\gamma_{\ell}\mu_{\ell}\rangle}\left[\frac{1}{M}\sum_{a}\mathrm{sgn}\left(\lambda_{a\mu_{1}}^{\gamma_{1}}u_{\mu_{1}}\right)\dots\mathrm{sgn}\left(\lambda_{a\mu_{\ell}}^{\gamma_{\ell}}u_{\mu_{\ell}}\right)\right]^{2}\right\} .\label{eq:upsi2}
\end{eqnarray}
Observe that $\Upsilon$ is the part of the replicated partition function
that accounts for the interaction between peers and the interaction
between peers and graph. If $\eta_{0}=\Delta=0$ then $\Upsilon=1.$
We define 
\begin{eqnarray}
\varrho_{\mu_{1}\dots\mu_{\ell}}^{\gamma_{1}\dots\gamma_{\ell}} & \equiv&\frac{1}{M}\sum_{a}\mathrm{sgn}\left(\lambda_{a\mu_{1}}^{\gamma_{1}}u_{\mu_{1}}\right)\dots\mathrm{sgn}\left(\lambda_{a\mu_{\ell}}^{\gamma_{\ell}}u_{\mu_{\ell}}\right)\label{eq:rho1}\\
\varrho_{0} & \equiv&\frac{1}{M}\sum_{a}1,\label{eq:rho0}
\end{eqnarray}
where $\varrho_{\mu_{1}\dots\mu_{\ell}}^{\gamma_{1}\dots\gamma_{\ell}}$
is the average level of agreement per individual, across issues and
replicas and $\varrho_{0},$ which is fancy way to write 1, will be
left as a free parameter for the time being until we apply a variational
technique (which will confirm its value, see below).

Applying Laplace's method to the integrals involving the parameters
$\varrho_{\mu_{1}\dots\mu_{\ell}}^{\gamma_{1}\dots\gamma_{\ell}}$
and their conjugates $\hat{\varrho}_{\mu_{1}\dots\mu_{\ell}}^{\gamma_{1}\dots\gamma_{\ell}}$
, we can express the replicated partition function as: 
\begin{eqnarray}
\overline{Z^{n}}(\beta) & =&\mathop{\mathrm{extr}}_{q,W,R,\left\{ \varrho_{\mu_{1}\dots\mu_{\ell}}^{\gamma_{1}\dots\gamma_{\ell}},\hat{\varrho}_{\mu_{1}\dots\mu_{\ell}}^{\gamma_{1}\dots\gamma_{\ell}}\right\} }\left\{ \exp\left[\frac{nNM}{2}\left(\ln(1-q)+\frac{q-R^{2}}{1-q}\right)\right]\right.\nonumber \\
 && \exp\left[-M\sum_{\ell=0}\sum_{\langle\gamma_{1}\mu_{1};\dots;\gamma_{\ell}\mu_{\ell}\rangle}\varrho_{\mu_{1}\dots\mu_{\ell}}^{\gamma_{1}\dots\gamma_{\ell}}\hat{\varrho}_{\mu_{1}\dots\mu_{\ell}}^{\gamma_{1}\dots\gamma_{\ell}}-\frac{\nu}{2}M+\right.\nonumber \\
 && \quad\left.+\frac{\nu}{2}M\sum_{\ell=0}\left\langle \cosh(\beta\eta)^{nP}\tanh(\beta\eta)^{\ell}\right\rangle _{\eta}\sum_{\langle\gamma_{1}\mu_{1};\dots;\gamma_{\ell}\mu_{\ell}\rangle}\left(\varrho_{\mu_{1}\dots\mu_{\ell}}^{\gamma_{1}\dots\gamma_{\ell}}\right)^{2}\right]\nonumber \\
 && \qquad\int\prod_{\mu}\mathcal{D}u_{\mu}\prod_{\mu}\mathcal{D}y_{\mu}\left(\prod_{\mu}\mathcal{D}t_{\mu}\prod_{\gamma\mu a}\frac{\mathrm{d}\lambda_{\mu}^{\gamma}}{\sqrt{2\pi}}\exp\left[-\frac{1}{2}\sum_{\gamma\mu}(\lambda_{\mu}^{\gamma}-\mathcal{C}_{\mu})^{2}+\right.\right.\nonumber \\
 && \qquad+\sum_{\ell=0}\sum_{\langle\gamma_{1}\mu_{1};\dots;\gamma_{\ell}\mu_{\ell}\rangle}\hat{\varrho}_{\mu_{1}\dots\mu_{\ell}}^{\gamma_{1}\dots\gamma_{\ell}}\mathrm{sgn}\left(\lambda_{\mu_{1}}^{\gamma_{1}}u_{\mu_{1}}\right)\dots\mathrm{sgn}\left(\lambda_{\mu_{\ell}}^{\gamma_{\ell}}u_{\mu_{\ell}}\right)+\nonumber \\
 && \quad\quad\left.\left.\left.+(1-\nu\eta_{0})\beta\sum_{\gamma\mu}\mathrm{sgn}\left(\lambda_{\mu_{1}}^{\gamma_{1}}u_{\mu_{1}}\right)\right]\right)^{M}\right\} \label{eq:z5}
\end{eqnarray}
where we have disregarded terms of $O(n^{2}),$ $O(N^{-1})$ and $O(M^{-1})$
in the argument of the exponential and now: 
\begin{equation}
\mathcal{C}_{\mu}\equiv\frac{\sqrt{W-R^{2}}y_{\mu}+Ru_{\mu}+\sqrt{q-W}t_{\mu}}{\sqrt{1-q}}.\label{eq:c1}
\end{equation}

Once more we consider the RS approach by introducing the distribution
$\pi(z)$ and its conjugate $\hat{\pi}(z):$

\begin{eqnarray}
\hat{\varrho}_{\mu_{1}\dots\mu_{\ell}}^{\gamma_{1}\dots\gamma_{\ell}}=\mathcal{C}_{\hat{\pi}}\int\mathrm{d}s\,\hat{\pi}(s)\tanh^{\ell}(\beta
s) & \qquad&\varrho_{\mu_{1}\dots\mu_{\ell}}^{\gamma_{1}\dots\gamma_{\ell}}=\int\mathrm{d}z\,\pi(z)\tanh^{\ell}(\beta z)\label{eq:pi1}\\
\hat{\varrho}_{0}=\mathcal{C}_{\hat{\pi}}\int\mathrm{d}s\,\hat{\pi}(s) & \qquad&\varrho_{0}=\int\mathrm{d}z\,\pi(z)\label{eq:pi2}
\end{eqnarray}
where equations (\ref{eq:pi1}) are the definitions of the fields
$\pi$ and $\hat{\pi}$, and equations $(\ref{eq:pi2})$ are the normalization
conditions they must satisfy. By using the symmetry of the RS supposition
and by integrating over the variables $\{\lambda_{\mu}^{\gamma}\}$,
the Hubbard-Stratanovich variables $u,$ $t$ and $y$ and by applying
the scaling condition $P=\alpha N,$ we have that the replicated partition
function takes the form of: 
\begin{eqnarray}
\overline{Z^{n}}(\beta) & =&\mathop{\mathrm{extr}}_{q,R,\varrho_{0},\hat{\varrho}_{0},\pi,\hat{\pi}}\left\{ \exp\left[-M\left(\frac{\nu}{2}-\hat{\varrho}_{0}+\varrho_{0}\hat{\varrho}_{0}-\frac{\nu}{2}\varrho_{0}^{2}\right)\right]\right.\nonumber \\
 & &\left(1+nMN\frac{\nu}{2}\varrho_{0}^{2}\alpha\left\langle \ln\cosh(\beta\eta)\right\rangle _{\eta}+\frac{nMN}{2}\left(\ln(1-q)+\frac{q-R^{2}}{1-q}\right)-\right.\nonumber \\
 && -nMN\alpha\mathcal{C}_{\hat{\pi}}\int\mathrm{d}z\,\mathrm{d}s\,\pi(z)\hat{\pi}(s)\ln\left[1+\tanh(\beta s)\tanh(\beta z)\right]+\nonumber \\
 && +\frac{\nu}{2}nMN\alpha\int\mathrm{d}z_{1}\,\mathrm{d}z_{2}\,\pi(z_{1})\pi(z_{2})\,\left\langle \ln\left(1+\tanh(\beta\eta)\tanh(\beta z_{1})\tanh(\beta z_{2})\right)\right\rangle _{\eta}+\nonumber \\
 && +nNM\alpha\mathrm{e}^{-\hat{\varrho}_{0}}\sum_{C=0}^{\infty}\frac{\mathcal{C}_{\hat{\pi}}^{\ell}}{C!}\int_{-\infty}^{\infty}\prod_{\ell=1}^{C}\mathrm{d}s_{\ell}\,\hat{\pi}(s_{\ell})\;2\int_{-\infty}^{\infty}\mathcal{D}x\mathcal{H}\left(-\frac{Rx}{\sqrt{q-R^{2}}}\right)\nonumber \\
 && \qquad\ln\left[\mathcal{H}\left(\sqrt{\frac{q}{1-q}}x\right)\mathrm{e}^{-\beta(1-\nu\eta_{0})}\prod_{\ell=1}^{C}[1-\tanh(\beta s_{\ell})]+\right.\nonumber \\
 && \qquad\quad\left.\left.\left.+\mathcal{H}\left(-\sqrt{\frac{q}{1-q}}x\right)\mathrm{e}^{\beta(1-\nu\eta_{0})}\prod_{\ell=1}^{C}[1+\tanh(\beta s_{\ell})]\right]\right)\right\} .\label{eq:z6}
\end{eqnarray}
Observe that during the integration process the dependency with respect
to $W$ disappears. By adding up the series, can be re-express as:
\begin{eqnarray}
\overline{Z^{n}}(\beta) & =&\mathop{\mathrm{extr}}_{q,R,\varrho_{0},\hat{\varrho}_{0},\pi,\hat{\pi}}\left\{ \exp\left[-M\left(\frac{\nu}{2}-\hat{\varrho}_{0}+\varrho_{0}\hat{\varrho}_{0}-\frac{\nu}{2}\varrho_{0}^{2}\right)\right]\right.\nonumber \\
 && \left(1+nMN\frac{\nu}{2}\varrho_{0}^{2}\alpha\left\langle \ln\cosh(\beta\eta)\right\rangle _{\eta}+\frac{nMN}{2}\left(\ln(1-q)+\frac{q-R^{2}}{1-q}\right)-\right.\nonumber \\
 && -nMN\alpha\mathcal{C}_{\hat{\pi}}\int\mathrm{d}z\,\mathrm{d}s\,\pi(z)\hat{\pi}(s)\ln\left[1+\tanh(\beta s)\tanh(\beta z)\right]+\nonumber \\
 && +\frac{\nu}{2}nMN\alpha\int\mathrm{d}z_{1}\,\mathrm{d}z_{2}\,\pi(z_{1})\pi(z_{2})\,\left\langle \ln\left(1+\tanh(\beta\eta)\tanh(\beta z_{1})\tanh(\beta z_{2})\right)\right\rangle _{\eta}+\nonumber \\
 && +nNM\alpha\left(-\beta(1-\nu\eta_{0})+\mathrm{e}^{-\hat{\varrho}_{0}}+\mathcal{C}_{\hat{\pi}}\int\mathrm{d}s\,\hat{\pi}(s)\ln[1+\tanh(-\beta s)]\right)\nonumber \\
 && +2nNM\alpha\int_{-\infty}^{\infty}\mathcal{D}x\mathcal{H}\left(-\frac{Rx}{\sqrt{q-R^{2}}}\right)\nonumber \\
 && \left.\left.\int\frac{\mathrm{d}y}{2\pi}\int\mathrm{d}\hat{y}\mathrm{e}^{-iy\hat{y}}\exp\left[\mathcal{C}_{\hat{\pi}}\int\mathrm{d}s\,\hat{\pi}(s)\mathrm{e}^{i\hat{y}s}-\hat{\varrho}_{0}\right]\ln\left[1+\left(\mathrm{e}^{2\beta(1-\nu\eta_{0}+y)}-1\right)\mathcal{H}\left(-\sqrt{\frac{q}{1-q}}x\right)\right]\right)\right\} .\label{eq:z7}
\end{eqnarray}
Observe that 
\begin{eqnarray}
\partial_{\varrho_{0}}\overline{Z^{n}} & =&\left(-M\hat{\varrho}_{0}+\nu M\varrho_{0}+O(n)\right)\overline{Z^{n}}\label{eq:rho00}\\
\partial_{\hat{\varrho}_{0}}\overline{Z^{n}} & =&\left(M-M\varrho_{0}+O(n)\right)\overline{Z^{n}}\label{eq:hatrho00}
\end{eqnarray}
which implies in the extreme that $\varrho_{0}=1$ (as it was expected
from equation (\ref{eq:rho0})) and $\hat{\varrho}_{0}=\nu$. From
this last equation we have that $\mathcal{C}_{\hat{\pi}}=\nu$. By
defining the weight function: 
\begin{equation}
\mathsf{P}(y|\hat{\pi})\equiv\int\frac{\mathrm{d}\hat{y}}{2\pi}\mathrm{e}^{-iy\hat{y}}\exp\left[\nu\left(\int\mathrm{d}s\,\hat{\pi}(s)\mathrm{e}^{i\hat{y}s}-1\right)\right],\label{eq:pfuncional}
\end{equation}
and the distribution 
\begin{equation}
\mathcal{P}(x|q,R)\equiv2\mathcal{N}(x)\mathcal{H}\left(-\frac{Rx}{\sqrt{q-R^{2}}}\right),\label{eq:distx}
\end{equation}
the free energy functional $F$ can be defined as 
\begin{eqnarray}
\beta F & \equiv& \frac{1-\overline{Z^{n}}(\beta)}{nNM}\nonumber \\
 & =& -\alpha\left(-\beta(1-\nu\eta_{0})+\mathrm{e}^{-\nu}+\frac{\nu}{2}\left\langle \ln\cosh(\beta\eta)\right\rangle _{\eta}\right)-\frac{1}{2}\left(\ln(1-q)+\frac{q-R^{2}}{1-q}\right)+\nonumber \\
 && -\frac{\nu}{2}\alpha\int\mathrm{d}z_{1}\,\mathrm{d}z_{2}\,\pi(z_{1})\pi(z_{2})\,\left\langle \ln\left(1+\tanh(\beta\eta)\tanh(\beta z_{1})\tanh(\beta z_{2})\right)\right\rangle _{\eta}+\nonumber \\
 && +\nu\alpha\int\mathrm{d}z\,\mathrm{d}s\,\pi(z)\hat{\pi}(s)\ln\left[\frac{1+\tanh(\beta s)\tanh(\beta z)}{1-\tanh(\beta s)}\right]-\alpha\beta\int\mathrm{d}y\mathsf{P}(y|\hat{\pi})(1-\nu\eta_{0}+y)\nonumber \\
 && -\alpha\int\mathrm{d}x\mathcal{P}_{Rq}(x)\int\mathrm{d}y\mathsf{P}(y|\hat{\pi})\ln\left[\mathrm{e}^{-\beta(1-\nu\eta_{0}+y)}\mathcal{H}\left(\sqrt{\frac{q}{1-q}}x\right)+\mathrm{e}^{\beta(1-\nu\eta_{0}+y)}\mathcal{H}\left(-\sqrt{\frac{q}{1-q}}x\right)\right] \label{eq:phi}
\end{eqnarray}
density of the system can be expressed as $f=\mathop{\mathrm{extr}}_{q,R,\pi,\hat{\pi}}F.$

\subsection{The characteristic function 
of the Distributions\label{sec:Map}}

By using the Fourier Transform in the saddle point equations (\ref{eq:sp1})
and (\ref{eq:sp2}) we define the functions: 
\begin{eqnarray}
\hat{\phi}(\omega) & \equiv&\int\mathrm{d}s\,\hat{\pi}(s)\mathrm{e}^{is\omega}\nonumber \\
 & =&1+i\mathcal{I}_{0}\omega-\frac{\mathcal{R}_{0}}{2}\omega^{2}+O(\omega^{3})\label{eq:reducedphipihat}\\
\phi(\omega) & \equiv&\int\mathrm{d}s\,\pi(s)\mathrm{e}^{is\omega}.\label{eq:phipi}
\end{eqnarray}
Thus:

\begin{eqnarray}
\mathsf{P}(y|\hat{\pi}) & \equiv&\int\frac{\mathrm{d}\omega}{2\pi}\,\mathrm{e}^{-iy\omega+\hat{\phi}(\omega)-\nu}\nonumber \\
 & \approx&\int\frac{\mathrm{d}\omega}{2\pi}\,\exp\left[-\frac{\nu\mathcal{R}_{0}}{2}\omega^{2}+i(\nu\mathcal{I}_{0}-y)\omega\right]=\mathcal{N}\left(y\left|\nu\mathcal{I}_{0},\nu\mathcal{R}_{0}\right.\right),\label{eq:pphihat}
\end{eqnarray}
which is consistent with (\ref{Uno}) and (\ref{Dos}). Consider the
definition (\ref{Lambda}), then the Fourier Transform of $\pi(s)$
can be expressed as: 
\begin{eqnarray}
\phi(\omega) & =&\int\mathrm{d}s\,\int\mathrm{d}y\,\mathsf{P}(y|\hat{\pi})\,\mathrm{e}^{is\omega}\left\langle \delta\left[s-1+\nu\eta_{0}-y-\frac{\Lambda(x)}{2}x^{2}\right]\right\rangle _{x}\nonumber \\
 & \approx&\int\mathrm{d}y\mathcal{N}\left(y\left|\nu\mathcal{I}_{0},\nu\mathcal{R}_{0}\right.\right)\mathrm{e}^{i\omega y}\left\langle \exp\left[i\omega\left(1-\nu\eta_{0}+\frac{\Lambda(x)}{2}x^{2})\right)\right]\right\rangle _{x},\label{eq:fi1}
\end{eqnarray}
where the expectation over $x$ is approximated by: 
\begin{eqnarray}
\left\langle \exp\left\{
i\omega\left[1-\nu\eta_{0}+\frac{\Lambda(x)}{2}\right]\right\} \right\rangle
_{x} & \approx&2\int_{-\infty}^{\infty}\mathcal{D}x\,\Theta(xR)\exp\left[i\omega\left(1-\nu\eta_{0}+\frac{\Lambda(x)}{2}x^{2})\right)\right]\nonumber \\
 & \approx&2\int_{0}^{\infty}\frac{\mathrm{d}x}{\sqrt{2\pi}}\exp\left[-\frac{x^{2}}{2}+i\omega\left(1-\nu\eta_{0}+\frac{\Lambda(R)}{2}x^{2}\right)\right]\nonumber \\
 & \approx&\int_{-\infty}^{\infty}\frac{\mathrm{d}x}{\sqrt{2\pi}}\exp\left[-[1-i\omega\Lambda(R)]\frac{x^{2}}{2}+is\left(1-\nu\eta_{0}\right)\right]\nonumber \\
 & \approx&\frac{1}{\sqrt{1-i\omega\Lambda}}\exp\left[i(1-\nu\eta_{0})\omega\right]\nonumber \\
 & \approx&\exp\left[-\frac{3}{8}\Lambda^{2}\omega^{2}+i\left(1-\nu\eta_{0}+\frac{\Lambda(R)}{2}\right)\omega\right],\label{eq:applam}
\end{eqnarray}
where (\ref{eq:applam}) is reached by assuming that $\Lambda$ sufficiently
small. From this point onwards we will refer to the parameter $\Lambda(R)$
as the amplified thermal noise. In such a case the function $\phi$
is Gaussian: 
\begin{equation}
\phi(s)\approx\exp\left[-\frac{\nu\mathcal{R}_{0}+\frac{3}{4}\Lambda(R)^{2}}{2}s^{2}+i\left(1-\nu\eta_{0}+\nu\mathcal{I}_{0}+\frac{\Lambda(R)}{2}\right)s\right],\label{eq:phiapp}
\end{equation}
and so 
\begin{equation}
\pi(z)\approx\frac{1}{\sqrt{2\pi\left(\nu\mathcal{R}_{0}+\frac{3}{4}\Lambda(R)^{2}\right)}}\exp\left\{ -\frac{\left(z-1+\nu\eta_{0}-\nu\mathcal{I}_{0}-\frac{1}{2}\Lambda(R)\right)^{2}}{2\left(\nu\mathcal{R}_{0}+\frac{3}{4}\Lambda(R)^{2}\right)}\right\} .\label{eq:apppi}
\end{equation}

The argument in the Dirac's delta function in (\ref{eq:sp2}) is mildly
sensitive to changes in the presidential approval, therefore we can approximate
it by: 
\begin{eqnarray}
\beta^{-1}\mathrm{arctanh}\left[\tanh(\beta\eta)\tanh(\beta z)\right] & \approx&\frac{|\eta+z|-|\eta-z|}{2}=\mathrm{sgn}(z\eta)\min\{|\eta|,|z|\},\label{eq:appT}
\end{eqnarray}
which allows us to write the expression of the Fourier transform of
the distribution $\hat{\pi}$ as: 
\begin{eqnarray}
\hat{\phi}(\omega) & =&\int\mathrm{d}s\phi(s)\left\langle \int\frac{\mathrm{d}z}{2\pi}\,\exp\left(-isz+i\omega\mathrm{sgn}(z\eta)\min\{|\eta|,|z|\}\right)\right\rangle _{\eta},\label{eq:fhatpi2}
\end{eqnarray}
where the expectation over the social strengths can be demonstrated
to be, disregarding terms of $O(\eta_{0}\Delta^{2})$: 
\begin{eqnarray}
\mathcal{Q} & \equiv&\left\langle \int\frac{\mathrm{d}z}{2\pi}\,\exp\left(-isz+i\omega\beta^{-1}\,\mathrm{arctanh}\left[\tanh(\beta\eta)\tanh(\beta z)\right]\right)\right\rangle _{\eta}\nonumber \\
 & \approx&\delta(s)+\frac{\omega}{\pi}\exp\left(-\frac{\Delta^{2}s^{2}}{2}\right)\frac{\eta_{0}}{s}-\omega^{2}\frac{\eta_{0}^{2}+\Delta^{2}}{2}\delta(s).\label{eq:Qexpand}
\end{eqnarray}
Expression (\ref{eq:Qexpand}), together with (\ref{eq:fhatpi2})
and (\ref{eq:reducedphipihat}) produce the following expressions
for the moments of $\nu^{-1}\hat{\pi}:$ 
\begin{eqnarray}
1 & =&\phi(0)\label{eq:m0}\\
\mathcal{I}_{0} & =&-\frac{i\eta_{0}}{\pi}\int\mathrm{d}s\frac{\phi(s)}{s}\mathrm{e}^{-\frac{\Delta^{2}s^{2}}{2}}\label{eq:m1}\\
\mathcal{R}_{0} & =&\eta_{0}^{2}+\Delta^{2},\label{eq:m2-1}
\end{eqnarray}
which implies that 
\begin{equation}
\mathcal{I}_{0}^{\star}=\eta_{0}\mathrm{erf}\left(\frac{1-\nu\eta_{0}+\nu\mathcal{I}_{0}^{\star}+\frac{1}{2}\Lambda(R)}{\sqrt{2\left[\nu(\eta_{0}^{2}+\Delta^{2})+\frac{3}{4}\Lambda(R)^{2}\right]}}\right).\label{eq:map}
\end{equation}
The equation (\ref{eq:map}) has one, two or three solutions depending
on the value of the parameters $\nu,$ $\eta_{0}$, $\Delta$ and
the function $\Lambda(R).$ It is easy to see that for the line: 
\begin{equation}
\eta_{0}=\frac{2+\Lambda(R)}{2\nu}\label{eq:cero}
\end{equation}
$\mathcal{I}_{0}^{\star}=0$ is a solution. Thus we expect that for
$\nu\eta_{0}>1+\Lambda(R)/2$, $\mathcal{I}_{0}^{\star}\approx-(\eta_{0}-\epsilon)$
is a solution, and for $\nu\eta_{0}<1+\Lambda(R)/2$, $\mathcal{I}_{0}\approx\eta_{0}-\epsilon$
is a solution (in both cases $0<\epsilon<\eta_{0}$ is a suitable
positive number).

If the derivative with respect to $\mathcal{I}_{0}$ of the right-hand-side
of (\ref{eq:map}) evaluated at $\mathcal{I}_{0}^{\star}$ is equal
to 1, and $\mathcal{I}_{0}^{\star}$ is also a solution of (\ref{eq:map}),
then (\ref{eq:map}) has two solutions, $\mathcal{I}_{0}^{\star}$
and $\eta_{0}-\epsilon$ or $-(\eta_{0}-\epsilon)$ depending on whether
$\nu\eta_{0}<1+\Lambda(R)/2$ or $\nu\eta_{0}>1+\Lambda(R)/2$, respectively.

If $\eta_{-}(\Lambda)<\eta_{0}<\eta_{+}(\Lambda),$ where: 
\begin{eqnarray}
\eta_{\pm} & =&\frac{2\mp|\Lambda|}{2\nu}\mp\left[\frac{1}{\nu}\sqrt{\left[\nu(\eta_{\pm}^{2}+\Delta^{2})+\frac{3}{4}\Lambda(R)^{2}\right]\log\left(\frac{2}{\pi}\frac{\nu^{2}\eta_{\pm}^{2}}{\nu(\eta_{\pm}^{2}+\Delta^{2})+\frac{3}{4}\Lambda(R)^{2}}\right)}-\right.\nonumber \\
 && \left.-\eta_{\pm}\mathrm{erf}\left(\sqrt{\log\left(\sqrt{\frac{2}{\pi}\frac{\nu^{2}\eta_{\pm}^{2}}{\nu(\eta_{\pm}^{2}+\Delta^{2})+\frac{3}{4}\Lambda(R)^{2}}}\right)}\right)\right]\label{eq:limits}
\end{eqnarray}
are the superior $(\eta_{+})$ and inferior $(\eta_{-})$ limits to
the area in the plane $(|\Lambda|,\eta_{0})$ where three solutions
to the equation (\ref{eq:map}) can be found. Let as define the set
$\mathbb{A}\coloneqq\{(|\Lambda|,\eta_{0}):\eta_{-}(\Lambda)<\eta_{0}<\eta_{+}(\Lambda)\}.$
(The use of $|\Lambda|$ instead of $\Lambda$ is due to the fact
that realizable solutions must satisfy $\mathrm{sgn}(R)=\mathrm{sgn}(1-\nu\eta_{0})$.
This point will be clarified when the equations involving $R$ and
$q$ are contemplated. See bellow.)

\begin{figure}
\begin{centering}
\includegraphics[scale=0.5]{fig5.eps} 
\par\end{centering}
\caption{Phase diagram of the system in terms of the parameters $\nu\eta_{0}$
and $|\Lambda|$. There are two phases separated by the line $\nu\eta_{0}=1$.
For $\nu\eta_{0}<1$ we have that $R>0$ and the average consensus
is in favor of $\bf B$. For $\nu\eta_{0}>1$ $R<0$ and the average position
of the agents is to form local alliances against the president $\bf B$.
In all the points of the plain outside the region $\mathbb{A}$, the
distribution describing the position of the neighborhood, given by
$\hat{\pi}(z)$, is sharply picked at $+\eta_{0}$, for the supportive
position, i.e. $\nu\eta_{0}<1$, or at $-\eta_{0}$, for the opposition
position, i.e. $\nu\eta_{0}>1$. In region $\mathbb{A}$ we have the
same phase separation at $\nu\eta_{0}=1$ but the contribution from
the neighborhood is a mixture of a opposition component plus a supportive
component. We also observe that the vertexes of $\mathbb{A}$ are
(1.651,0), (1,0.411) and (0.717,0).}
\label{mapa-2} 
\end{figure}
It is important to note that for very high values of the presidential approval
$\beta$ we have that:

\begin{equation}
\nu\eta_{\pm}\approx\frac{1}{1\pm\sqrt{\frac{1}{\nu}\log\left(\frac{2\nu}{\pi}\right)}\mp\mathrm{erf}\left(\sqrt{\frac{1}{2}\log\left(\frac{2\nu}{\pi}\right)}\right)},\label{limites}
\end{equation}
which implies that the segment (at very high presidential approval) of the coexistence
of states grows with $\sqrt{\nu}$. From equation (\ref{eq:map})
we conclude that, to satisfy the saddle-point equations (\ref{eq:sp1},\ref{eq:sp2}),
there must be a set of values of $\eta$ such that, for a very high
presidential approval $\beta$ there coexist states with different attitudes.

For all the points $(\Lambda,\eta_{0})\notin\mathbb{A},$ we have
that both conditions (\ref{eq:m1}) and (\ref{eq:m2-1}) are satisfied
for a distribution: 
\begin{eqnarray}
\hat{\pi}(s) & =&\mathcal{N}\left(s\left|\mathcal{I}_{0}^{\star},\eta_{0}^{2}-(\mathcal{I}_{0}^{\star})^{2}+\Delta^{2}\right.\right)\label{eq:pihatsola}
\end{eqnarray}
where $\mathcal{I}_{0}^{\star}$ is the only solution of (\ref{eq:map}).
In this case we also have that: 
\begin{equation}
\mathsf{P}(y|\hat{\pi})\approx\mathcal{N}\left(y\left|\nu\mathcal{I}_{0}^{\star},\nu(\eta_{0}^{2}+\Delta^{2})\right.\right).\label{eq:fouriersola}
\end{equation}

For the points $(\Lambda,\eta_{0})\in\mathbb{A}$ we propose the following
form for $\hat{\pi}(s)=\mathcal{Z}^{-1}\exp[-\Phi(s)],$ where $\mathcal{Z}$
is a normalization constant and the function $\Phi(s)$ is defined
as:

\begin{eqnarray}
\Phi(s) & \coloneqq&\frac{s^{2}}{2\Delta^{2}}-\frac{\eta_{0}}{\nu}\frac{1-\nu\eta_{0}+\nu s+\frac{1}{2}\Lambda(R)}{\Delta^{2}}\mathrm{erf}\left(\frac{1-\nu\eta_{0}+\nu s+\frac{1}{2}\Lambda(R)}{\sqrt{2\left[\nu(\eta_{0}^{2}+\Delta^{2})+\frac{3}{4}\Lambda(R)^{2}\right]}}\right)-\nonumber \\
 && -2\frac{\eta_{0}}{\nu}\frac{\nu(\eta_{0}^{2}+\Delta^{2})+\frac{3}{4}\Lambda(R)^{2}}{\Delta^{2}}\mathcal{N}\left(\nu s\left|\nu\eta_{0}-1-\frac{1}{2}\Lambda(R),\nu(\eta_{0}^{2}+\Delta^{2})+\frac{3}{4}\Lambda(R)^{2}\right.\right).\label{Phi-1}
\end{eqnarray}
Observe that $\Phi'(\mathcal{I}_{0}^{\star})=0$ is identical to (\ref{eq:map})
and, for all points in $\mathbb{A}$ this equations have three roots
$\mathcal{I}_{-}^{\star}<\mathcal{I}_{0}^{\star}<\mathcal{I}_{+}^{\star}.$
The asymptotic behavior of $\lim_{s\to\pm\infty}s^{-2}\Phi(s)>0$
indicates that $\mathcal{I}_{\pm}^{\star}$ are minima with $\mathcal{I}_{0}^{\star}$
an intermediate maximum. Let us compute the second derivative of $\Phi$
at the minima: 
\begin{eqnarray}
\Phi''(\mathcal{I}_{\pm}^{\star}) & =&\frac{1}{\Delta^{2}}\left(1-\eta_{0}\left.\frac{\mathrm{d}}{\mathrm{d}s}\mathrm{erf}\left(\frac{1-\nu\eta_{0}+\nu s+\frac{1}{2}\Lambda(R)}{\sqrt{2\left[\nu(\eta_{0}^{2}+\Delta^{2})+\frac{3}{4}\Lambda(R)^{2}\right]}}\right)\right|_{s=\mathcal{I}_{\pm}^{\star}}\right),\label{eq:segunda}
\end{eqnarray}
therefore $\Phi''(\mathcal{I}_{\pm}^{\star})=\mathcal{L}_{\pm}^{2}\Delta^{-2},$
with: 
\begin{equation}
\mathcal{L}_{\pm}\coloneqq\sqrt{1-2\nu\eta_{0}\mathcal{N}\left(\nu\mathcal{I}_{\pm}^{\star}\left|\nu\eta_{0}-1-\frac{1}{2}\Lambda(R),\nu(\eta_{0}^{2}+\Delta^{2})+\frac{3}{4}\Lambda(R)^{2}\right.\right),}\label{eq:Lambda}
\end{equation}
which is larger than zero for all $(\Lambda,\eta_{0})\in\mathbb{A}$.
Thus: 
\begin{eqnarray}
\hat{\pi}(s) & \approx & h_{+}\mathcal{N}\left(s\left|\mathcal{I}_{+}^{\star},\Delta^{2}\right.\right)+h_{-}\mathcal{N}\left(s\left|\mathcal{I}_{-}^{\star},\Delta^{2}\right.\right)\label{eq:mezcla}\\
h_{\pm} & \equiv & \frac{1}{2}\pm\frac{1}{2}\tanh\left(\frac{\Phi(\mathcal{I}_{+}^{\star})-\Phi(\mathcal{I}_{-}^{\star})}{4}-\frac{1}{2}\ln\frac{\mathcal{L}_{+}}{\mathcal{L}_{-}}\right)\label{eq:hache}\\
\int\mathrm{d}s\hat{\pi}(s)s & \approx & h_{+}\mathcal{I}_{+}^{\star}+h_{-}\mathcal{I}_{-}^{\star}\label{eq:mezcla1}\\
\int\mathrm{d}s\hat{\pi}(s)s^{2} & \approx & h_{+}(\mathcal{I}_{+}^{\star})^{2}+h_{-}(\mathcal{I}_{-}^{\star})^{2}+\Delta^{2},\label{eq:var}
\end{eqnarray}
observe that the expectation (104) represents a convex combination
between the solutions to the equation (\ref{eq:map}), and $\hat{\pi}'(\mathcal{I}_{\pm}^{\star})\approx0.$
In order to compute the distribution $\mathsf{P}(y|\hat{\pi})$ we
first need to compute the Fourier transform of $\hat{\pi}$: 
\begin{eqnarray}
\hat{\phi}(\omega) & \approx& h_{+}\exp\left[-\frac{\Delta^{2}}{2}\omega^{2}+i\omega\mathcal{I}_{+}^{\star}\right]+h_{-}\exp\left[-\frac{\Delta^{2}}{2}\omega^{2}+i\omega\mathcal{I}_{-}^{\star}\right],\label{eq:fhat}
\end{eqnarray}
and if we define 
\begin{eqnarray}
\Upsilon(\omega,y) & \coloneqq&-iy\omega-\nu+\nu\int\mathrm{d}s\frac{\exp[-\Phi(s)+is\omega]}{\int\mathrm{d}s'\,\exp[-\Phi(s')]}\nonumber \\
 & \approx&-iy\omega-\nu+\nu\left\{ h_{+}\exp\left[-\frac{\Delta^{2}}{2}\omega^{2}+i\omega\mathcal{I}_{+}^{\star}\right]+h_{-}\exp\left[-\frac{\Delta^{2}}{2}\omega^{2}+i\omega\mathcal{I}_{-}^{\star}\right]\right\} \nonumber \\
 & \approx&-iy\omega+\nu\left\{ i\omega(h_{+}\mathcal{I}_{+}^{\star}+h_{-}\mathcal{I}_{-}^{\star})-\frac{h_{+}(\mathcal{I}_{+}^{\star})^{2}+h_{-}(\mathcal{I}_{-}^{\star})^{2}+\Delta^{2}}{2}\omega^{2}\right\} +O(\omega^{3})\label{eq:upsi-1}
\end{eqnarray}

\begin{eqnarray}
\mathsf{P}(y|\hat{\pi}) & =&\int\frac{\mathrm{d}\omega}{2\pi}\,\mathrm{e}^{\Upsilon(\omega,y)}\nonumber \\
 & \approx&\mathcal{N}\left(y\left|\nu(h_{+}\mathcal{I}_{+}^{\star}+h_{-}\mathcal{I}_{-}^{\star}),\nu[h_{+}(\mathcal{I}_{+}^{\star})^{2}+h_{-}(\mathcal{I}_{-}^{\star})^{2}+\Delta^{2}]\right.\right)\label{eq:dos}
\end{eqnarray}

These distributions should be applied depending on the value of $\eta_{0}$
and $\Lambda(R),$ following the diagram presented in figure \ref{mapa-2}.


\end{document}